\documentclass[11pt]{article}

\usepackage{graphicx}
\usepackage{latexsym}
\usepackage{amssymb}
\usepackage{amsmath}

\numberwithin{equation}{section}
\def\tr{{\rm tr}}
\def\diag{{\rm diag}\,}
\def\Tr{{\rm Tr}}

\def\exp{{\rm exp}\,}

\def\S{{\bf S}}
\def\C{{\bf C}}
\def\Z{{\bf Z}}
\def\R{{\bf R}}

\def\P{{\bf P}}
\def\T{{\bf T}}

\def\CF{{\cal F}}

\def\CO{{\cal O}}

\def\CX{{\cal X}}
\def\CY{{\cal Y}}
\def\CW{{\cal W}}

\def\centeron#1#2{{\setbox0=\hbox{#1}\setbox1=\hbox{#2}\ifdim

   \wd1>\wd0\kern.48\wd1\kern-.48\wd0\fi

   \copy0\kern-.48\wd0\kern-.48\wd1\copy1\ifdim\wd0>\wd1

   \kern.48\wd0\kern-.48\wd1\fi}}

\newcommand{\beq}{\begin{equation}}

\newcommand{\eeq}{\end{equation}}

\newcommand{\bea}{\begin{eqnarray}}

\newcommand{\eea}{\end{eqnarray}}

\newcommand{\ba}{\begin{array}}

\newcommand{\ea}{\end{array}}

\newcommand{\p}{\partial}

\newcommand{\nn}{\nonumber}

\newcommand{\half}{\frac{1}{2}}

\begin{document}

\hskip5cm

\centerline{\LARGE \bf Non-compact Topological Branes on Conifold}

\vskip2cm

\hskip2in

\centerline{\Large Seungjoon Hyun and Sang-Heon Yi}

\vskip0.5cm

\centerline{\it Department of Physics, College of Science, Yonsei University, Seoul 120-749, Korea}

\vskip2cm

\centerline{\bf Abstract} We consider non-compact branes in
topological string theories on a class of Calabi-Yau spaces
including the resolved conifold and its mirror. We compute the
amplitudes of the insertion of non-compact Lagrangian branes in the
A-model on the resolved conifold in the context of the topological
vertex as well as the melting crystal picture. They all agree with each other and also agree with the
results from Chern-Simons theory,  supporting the large $N$ duality.  We find that they obey the Schr\"odinger equation confirming the wavefunction behavior of the amplitudes. We also compute the amplitudes of the
non-compact B-branes in the DV matrix model which arises as a
B-model open string field theory on the mirror manifold of the
deformed conifold. We take the large $N$ duality to consider the
B-model on the mirror of the resolved conifold and
confirm the wave function behavior of this amplitude. We find
appropriate descriptions of non-compact branes in each model, which
give complete agreements among those amplitudes and clarify the
salient features including the role of symmetries toward these
agreements.

\hskip3cm

\centerline{E-mail : hyun@phya.yonsei.ac.kr,
shyi@phya.yonsei.ac.kr}
\thispagestyle{empty}

\newpage
\section{Introduction}
Topological strings have been a very interesting test ground for
various dualities. These dualities in topological strings give us
various relations among mathematically seemingly unrelated
objects. One of the well-known examples is the mirror symmetry
between the two Calabi-Yau threefolds.
 A-model topological string theory on a Calabi-Yau space is equivalent
 to B-model topological string theory on the mirror Calabi-Yau space
 and vice versa(see, for a
review~\cite{Hori:2003ic}\cite{Marino:2004uf}\cite{Marino:2004eq}\cite{Neitzke:2004ni}).

 Another relation found recently is
the large $N$ duality via geometric transition, which is a
topological incarnation of the open-closed duality or AdS/CFT
correspondence. The A-model open string field theory on the Calabi-Yau
spaces of the form $T^*M$ with $N$ Lagrangian branes wrapped on
$M$ reduces to the $U(N)$ Chern-Simons theory on
$M$~\cite{Witten:1992fb}.  In particular, Chern-Simons theory on
$\S^3$ describes the A-model open string theory on the deformed
conifold, $T^*\S^3$. Through the geometric transition, it is found
to be equivalent to the A-model on the resolved conifold
\cite{Gopakumar:1998ki}\cite{Berkovits:2003pq}.

Traditionally the exact computaion of correlation functions in the
topological A-model was regarded as a difficult one because of the
worldsheet instanton contributions. This has been changed greatly
at least for the model on non-compact toric Calabi-Yau spaces, by
introducing topological vertex~\cite{Aganagic:2003db}. Main idea
is that the toric Calabi-Yau space can be regarded as the
composition of $\C^3$'s. More concretely, the toric $\C^3$ can be
represented by a toric diagram as a trivalent vertex, and toric
Calabi-Yau spaces can be constructed diagrammatically by gluing
these vertices appropriately. The A-model topological string
amplitudes on toric Calabi-Yau spaces can be obtained likewise
from the topological vertex, which is the open string amplitude on
$\C^3$ for non-compact A-branes at three legs in the toric diagram
of $\C^3$. Along these lines, the vacuum partition function of the
A-model on the resolved conifold was reproduced. The computation
can be easily generalized to the case when the background contains
non-compact Lagrangian A-branes.

Another interesting progress in the A-model has been made from the
observation that the partition function of the topological A model
on $\C^3$ is the same as the partition function of
three-dimensional melting crystal filling on the positive octant
of $\R^3$~\cite{Okounkov:2003sp}. Again this relation seems to be
generalized to some toric Calabi-Yau spaces. Indeed the vacuum
partition function of the topological A model on the resolved
conifold was shown to be reproduced by considering melting cristal
model with an extra wall in one direction~\cite{Okuda:2004mb}. In
this melting crystal model, defects play the role of the
Lagrangian A-branes. Some computations along this line has been
made in~\cite{Saulina:2004da}.

On the other hand, remarkable progress on the computation of the
correlation function in the topological B-model has been made as
well \cite{Aganagic:2003qj}. These developments came from the
realization of the $\CW$-algebra governing the correlation
functions of the topological B-model on the class of local
Calabi-Yau threefolds which is given by
\bea uv= H(x,y)~. \label{CY} \eea
The symmetry comes from the holomorphic diffeomorphism of the
Calabi-Yau spaces which preserves the holomorphic three form. All
the relevant features of the B-model in this class of local
Calabi-Yau threefolds can be obtained from the Riemann surface
which satisfies \bea H(x, y)=0~. \eea In this model the
non-compact B-brane is identified as a fermion in the Riemann
surface. 

There is a large $N$ duality in the B-model as well. The B-model
on a Calabi-Yau space $\widetilde{\CY}$ with the flux on $S^3$
cycle is equivalent to the open B-model on $N$ holomorphic branes
wrapped on $\P^1$ in the Calabi-Yau space $\CY$ which is related
to $\widetilde{\CY}$ via geometric transition. The B-model open
string field theory on $\CY$ becomes Dijkgraaf-Vafa(DV) matrix
model \cite{Dijkgraaf:2002fc}.

All these relations, including the mirror symmetry and the large $N$
transition, among various theories are expected to hold for more
general background. Especially, the non-compact branes should play an
essential role to establish the full equivalence of those
theories. For instance, in the Chern-Simons theory, viewed as an open
string field theory, the insertion of the non-compact branes gives the
generating functional of the correlation functions of gauge invariant
observables, namely knot invariants. It plays the same role in the DV
matrix model.  In the context of the B-model closed strings, the
insertion of non-compact B-brane generates the deformations of complex
structure. All these indicate that it is essential to include
non-compact B-branes to establish the full equivalence among these
theories.
However, there seems to be some subtleties in the
relations among these theories in the presence of
non-compact branes. Those subtleties have often been overlooked, without causing  problems, in the simple model like topological vertex on $\C^3$ and the B-model on the deformed conifold.

In this paper, we revisit the topological
string theories with
non-compact branes. We consider the A-model  strings on the resolved
conifold and the B-model on its mirror manifold as well as the open strings on their large $N$ dual
Calabi-Yau spaces. In particular we use the topological vertex formalism to compute the amplitudes of the non-compact Lagrangian brane at the various legs in the toric diagram of the resolved conifold.
We also compute the amplitudes of the defect in the melting crystal model. In the B-model,
we compute the amplitude of non-compact holomorphic brane using the DV matrix model and
use the large $N$ duality to identify it as the amplitude of non-compact brane in the B-model on the mirror of the resolved conifold. In all these computations, along with the known results in Chern-Simons theory, we find the correct identification
among various parameters and quantities in these models. We also find how the underlying $SL(2\,, \Z)$
symmetries are realized and related among these models with
non-compact branes. These computations show manifestly the wavefunction behavior of the A-model amplitude, which seems to be a natural consequence of the mirror symmetry. All these show clearly how the non-compact branes in various models are implemented and how they are connected through the chain of mirror symmetry and the large $N$ duality in the topological string theories.
Very recently, a paper \cite{Kashani-Poor:2006nc} appeared which deals with the 
amplitudes of non-compact branes in the B-model on the mirror of the
resolved conifold using the fermionic formulation in \cite{Aganagic:2003qj}.  
The results in the present paper along with their work  
complete all those dualities in the presence of the non-compact branes on the class of toric Calabi-Yau spaces.

The organization of
this paper is as follows: In the next section,
we review various topological strings theories and related models,
partly in order to establish the notations and the relations among
various parameters in those theories. We also review non-compact branes in the open string field theory on $\S^3$ in the deformed conifold, which play the role of sources in the generating functional of knot invariants in the $U(N)$ Chern-Simons theory on $\S^3$.  In section 3, we describe
topological vertex and compute the amplitudes of the non-compact branes at the various legs in the toric diagram of the resolved conifold. In section 4, we describe  the melting crystal models and compute the amplitudes of  the defects and show that they correspond to the non-compact branes in topological vertex as expected.
In section 5, we obtain the B-model amplitudes of non-compact branes in the mirror of the resolved  conifold from the large $N$ dual DV matrix model. In section 6,  we show that all these amplitudes in various models are equivalent obeying the same differential equations as well as the same transformation rules under $SL(2\,, \Z)$. In particular we show the wavefunction behavior of the A-model amplitudes confirming the suggestion from the mirror symmetry.  Finally we draw some conclusions.

\section{Topological string theories and their duals}

In this section we review the mirror symmetry which relates the
A-model topological string theory on the resolved conifold and the
B-model on its mirror Calabi-Yau space. We also review the large
$N$ duality which connects the closed topological strings on a
Calabi-Yau manifold with the open topological strings on another
Calabi-Yau space which is related to the former via the geometric
transition. This large $N$ duality relates the A-model topological
string theory on the resolved conifold with $U(N)$ Chern-Simons
theory on $\S^3$, which is an open string field theory on the
deformed conifold. We also consider  the amplitude of non-compact
branes in the open topological string theories on the deformed
conifold. In $U(N)$ Chern-Simons theory on $\S^3$, which is the
open topological A model on deformed conifold, a non-compact brane appears as a
Wilson loop and gives rise to the generating functional of knot
invariants.

\subsection{The large $N$ duality and the mirror symmetry }

In this subsection we explain the large $N$ duality which relates open
and closed topological strings on Calabi-Yau(CY) threefolds and the
mirror symmetry which relates the A- and B-models.

The large $N$ duality in the A-model topological string theory means that two
CY threefold ${\cal X}$ and $\widetilde{\cal X}$ are
related by the geometric transition while physics on two CY's is
identical.  As the three cycle $\S^3$ wrapped by a large number $N$
of A-branes in ${\cal X}$ shrinks down, ${\cal X}$ becomes a
singular CY threefold,  and then by blowing up the singular point,
two cycle $\P^1$ with RR-flux appears while A-branes on $\S^3$ disappear. As a
result of this geometric transition, ${\cal X}$ becomes another non-singular CY threefold
$\widetilde{\cal X}$. The claim is that in the large $N$ limit, open
topological string theory on A-branes is identical to the closed
topological string theory on $\widetilde{\cal X}$.

The deformed conifold ${\cal X}=T^{*}\S^3$  can be described by the
complex equation as
\beq {\cal X}~; \qquad uv = H(x,y) = xy-\mu\,, \eeq
where $\mu$ may be understood as the complex structure moduli and
corresponds to the size of three cycle $\S^3$.  We consider
A-model topological open strings on  $N$ A-branes
wrapped on a Lagrangian submanifold $\S^3$.  Then the open topological
string field theory on A-branes reduces to the $U(N)$ Chern-Simons
theory on $\S^3$.

\begin{figure}
\begin{center}
\includegraphics[width=8cm,height=6cm]{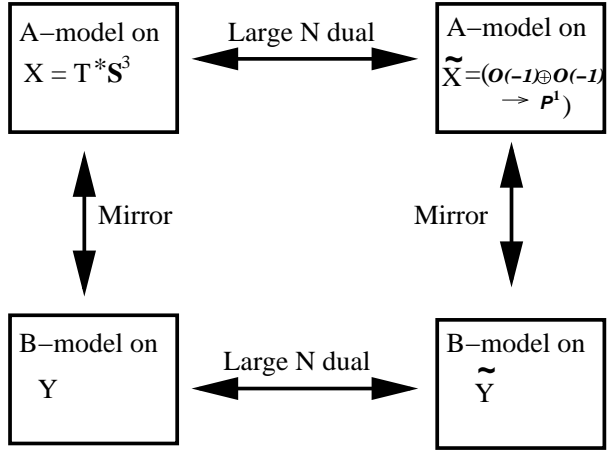}
\caption{Relations among various models} \label{default1}
\end{center}
\end{figure}

According to the
geometric transition we may set $\mu=0$ and get $\widetilde{\cal
X}$ by blowing up the singular geometry as
\beq \widetilde{\cal X}~; \qquad uz = x\,, \qquad vz' = y \,, \eeq
where $z$ and $z'$ are inhomogeneous coordinates on $\P^1$ and
are related by $z'=1/z$. It turns out that $\widetilde{\cal X}$ is the
resolved conifold, $\CO(-1)\oplus \CO(-1) \rightarrow\P^1$, which is
the total space of rank two vector bundle over $\P^1$.
$N$ A-branes on $\S^3$ are replaced by the RR-flux on $\P^1$ through the geometric transition.
According to the large $N$ duality, the closed topological A-model on this resolved
conifold is equivalent to the $U(N)$ Chern-Simons theory on
$\S^3$. In the closed string side we have two parameters, the string
coupling $g_s$ and the K\"{a}hler parameter $t$, while in the gauge
theory side we also have two parameters, the gauge coupling $g_{CS}^2$
and the rank of gauge group $N$.  In this large $N$ duality,  they
should be identified as
\beq g_s = ig_{CS}^2  \equiv i\frac{2\pi}{k+N}\,, \qquad t= g_s N, \eeq
where $g^2_{CS}$ is the coupling constant with the finite
renormalization effect.  Indeed it was shown in
\cite{Gopakumar:1998ki} that the vacuum partition
function of $U(N)$ Chern-Simons theory can be resummed using
$\frac{1}{N}$ expansion and shown to be identical with the vacuum
partition function of the topological A-model on the resolved
conifold\footnote{The presence of RR flux doesn't
  change the partition function of the model\cite{Vafa:2000wi}.}.

The CY threefold $\widetilde{\cal Y}$ mirror to the resolved
conifold $\widetilde{\cal X}$ can be obtained via a mirror map as
shown by Hori and Vafa\cite{Hori:2000kt}. For this purpose let us
recall the gauged linear sigma model(GLSM) or symplectic quotient
description of the resolved conifold~\cite{Witten:1993yc}
\beq
\label{resolved}
\widetilde{\cal X}~;\qquad \Big(|\phi_1|^2 + |\phi_2|^2
-|\phi_3|^2 -|\phi_4|^2 = {\rm Re}~t~\Big) /U(1)\,, \eeq
 where $U(1)$ group acts on the coordinates as
\bea
(\phi_1,\phi_2, \phi_3, \phi_4) \rightarrow (e^{i\alpha}\phi_1,
e^{i\alpha}\phi_2, e^{-i\alpha}\phi_3, e^{-i\alpha}\phi_4)\,.
\eea
Variables between two descriptions of the resolved conifold, $\widetilde{\CX}$, are related by
\beq x = \phi_1\phi_3\,, \qquad y = \phi_2\phi_4\,, \qquad u =\phi_2\phi_3 \,, \qquad v =\phi_1\phi_4\,, \qquad \quad ; \quad \qquad z=\frac{1}{z'}=\phi_1/\phi_2\,.
\eeq
%

\begin{figure}
\begin{center}
\includegraphics[width=8cm,height=6cm]{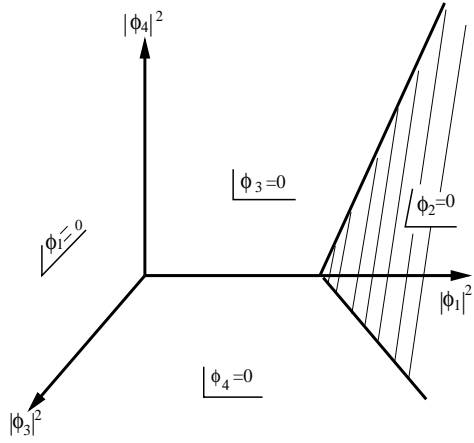}
\caption{Toric base for the resolved conifold} \label{toricbase}
\end{center}
\end{figure}

The explicit mirror map is given by
\beq
{\rm Re}\,Y_i = |\phi_i|,\qquad y_i =e^{-Y_i}
\eeq
and a Landau-Ginzberg superpotential which characterizes the mirror
manifold $\widetilde{\cal Y}$ is given by
\begin{equation}
H(y_i) = \sum_{i}y_i\,, \qquad y_1y_2 = e^{-t}y_3y_4\,,
\label{LG}
\end{equation}
Note
that $y_i$'s are $\C^{*}$ variables and the second equation is a constraint
 coming from GLSM.

One may use the inhomogeneous coordinates to express this curve equation. For example, one may eliminate $y_2$ using the constraint equation and use $\C^*$ action to set  $y_1=1$. Introducing variables $x,y$ as $y_3=-e^{-x},
y_4=-e^{y}$, one get the  local CY threefold $\widetilde{\cal Y}$
governed by the complex equation
\beq uv = H(x,y) =1- e^{y} - e^{-x} + e^{-t-x+y} \,,
\label{LG1}\eeq
on which topological B-model strings are
equivalent to the topological A-model strings on the resolved
conifold $\widetilde{\cal X}$.

Since a linear $SL(2\,,\,\Z)$ coordinate transformation does not
alter the periodicity of $x,y$ and the holomorphic three-form
$\Omega = dxdydv/v$, the mirror manifold $\widetilde{\cal Y}$ can
be alternatively described in terms of new variables ${\tilde x}$ and ${\tilde y}$
\beq
{{\tilde x} \choose
  {\tilde y}} = \left( \ba{rr} 1 & -1 \\ 0& 1 \ea\right) {x \choose y} + {-t \choose 0}\,,
\eeq
with a constant rescaling
$e^{x+y+\pi i}uv={\tilde u}{\tilde v}$ as
\beq \label{mirror}\widetilde{\cal Y}~; \qquad  {\tilde u}{\tilde v} = H({\tilde x},{\tilde y}) =
(e^{\tilde y}-1)(e^{{\tilde y}+{\tilde x}}-1)- \tilde{\mu}\,, \qquad \tilde{\mu} \equiv
1-e^t\,, \eeq
where $\tilde{\mu}$ can be interpreted as the complex structure
moduli. Note that this geometry is non-singular for
$\tilde{\mu}\neq 0$ and contains three cycles.

In this B-model
setting one may also consider analogous large $N$ dual B-model
open topological strings as in the A-model case, which can be achieved by a
geometric transition. Consider the topological B-model on the above
Calabi-Yau space with fluxes on $\S^3$. Now one let the complex
structure parameter $\tilde{\mu}$ go to zero and the corresponding
three cycle shrinks. Then, new two cycle appears and $N$ B-branes
wrapped on this two cycle are dual to the flux on the shrunken
three cycle. The manifold ${\cal Y}$ relevant for this large $N$
dualized B-model strings, which is also mirror to the starting
geometry ${\cal X}$, is given by the blow-up of the local CY
threefold $\widetilde{\cal Y}$ with $\tilde{\mu}=0$ as
\beq {\cal Y}~;\qquad  {\tilde u}z= e^{\tilde y} -1\,, \qquad {\tilde v}z' = e^{{\tilde y}+{\tilde x}}-1\,,
\eeq
where $z$ and $z'$ denote the inhomogeneous coordinates for the
blown up $\P^1$ and are related to each other by $z'=1/z$. The open string field
theory on B-branes wrapped on $\P^1$ reduces to the
Dijkgraaf-Vafa matrix model~\cite{Aganagic:2002wv}.

\subsection{The A-model on the resolved conifold}

In this section we summarize some relevant results and establish the
notations on the A-model topological string theory on the resolved
conifold for later use.
In general the vacuum partition function of topological closed strings
on a Calabi-Yau manifold\footnote{The topological A-model makes sense on
  K\"{a}hler manifolds.}
may be written as
\beq {\cal Z}(g_s,t) \equiv \exp[{\cal F}(g_s,t)] = \exp
\bigg[\sum_{g=0}^{\infty}g_s^{2g-2}F_g(t)\bigg]\,, \eeq
where $t$ denotes K\"{a}hler and complex structure moduli for
A-model and B-model topological string theory, respectively, and $F_g(t)$ is the genus-$g$ free
energy. The topological A-model closed string
amplitude on the resolved conifold was obtained by embedding in
physical M theory \cite{Gopakumar:1998ii}. It is given
by\footnote{Note that
there is an ambiguity in the expression for $g=0,1$ coming from the
non-compactness of the given CY.}
\beq \label{part-conifold} {\cal F}(g_s,t) = {\CF}^{const.}(g_s) +
\sum_{n=1}^{\infty}\frac{1}{n}\frac{e^{-nt}}{(2\sin\frac{ng_s}{2})^2}\,,
\eeq
where ${\cal F}^{const.}(g_s)$  denotes the constant map
contribution and $t$ is the K\"{a}hler modulus of $\P^1$. The
genus-$g$ free energy for the resolved conifold can be read as
\beq
F_g(t)
=
\frac{|2g-1||B_{2g}|}{(2g)!}\sum_{n=1}^{\infty}\frac{e^{-nt}}{n^{3-2g}}+
F^{const.}_{g}\,.   \eeq
{}From now on we adopt different conventions and take  the partition
function of A-model topological
strings on the resolved conifold as
\beq \label{part}
Z_{\widetilde{\CX}}(g_s,t) =
\exp\Big[F_{\widetilde{\CX}}(g_s,t)\Big] =
\exp\bigg[F^{const.}(g_s) -
\sum_{n=1}^{\infty}\frac{1}{n}\frac{e^{-nt}}{(2\sinh\frac{ng_s}{2})^2}\bigg]\,.
\eeq
One may regard this convention is coming from the replacement of  the
weight $g_s^{2g-2}$ by another weight $(-ig_s)^{2g-2}$ such that
\beq Z(g_s,t) \equiv \exp[F(g_s,t)] = \exp\bigg[
\sum_{g=0}^{\infty}(-ig_s)^{2g-2}F_g(t)\bigg]\,. \eeq
The explicit form of the constant map contribution on the resolved
conifold, $F^{const.}(g_s)$, is
\beq F^{const.}(g_s) =
\sum_{n=1}^{\infty}\frac{1}{n}\frac{1}{(2\sinh\frac{ng_s}{2})^2} =
\ln M(q_c)\,, \qquad M(q_c)\equiv
\prod_{m=1}^{\infty}(1-q^m_c)^{-m}\,, \quad q_c\equiv
e^{-g_s}=q^{-1}\,, \label{constant}\eeq
where $M(q_c)$ is the, so-called, Mac-Mahon function with ${\rm
  Re}~g_s > 0$. In the $t\rightarrow \infty$ limit, the partition
function on resolved conifold  reduces to the Mac-Mahon function, thus
which is the partition function of topological A-model closed strings
on $\C^3$. For the genus
$g\ge 2$, the genus-$g$ free energy may also be written as
\beq
F^{const.}_{g}=(-1)^g\frac{|B_{2g}B_{2g-2}|}{2g(2g-2)(2g-2)!}=(-1)^g\frac{\chi(\widetilde{\CX})}{2}
\int_{\overline{\cal M}_{g}} c^3_{g-1}({\cal H}) \,, \qquad g \ge
2\,, \eeq
where $B_{2g}$ is a Bernoulli number and $\chi(\widetilde{\CX})$ is
the Euler characteristic of the CY manifold $\widetilde{\CX}$.

\subsection{Non-compact branes in the open topological A model}

The amplitude of non-compact branes can be easily incorporated in the Chern-Simons theory.
Our starting set up is A-model topological open strings on the deformed
conifold ${\cal X}=T^{*}\S^3$ with a large number $N$ of A-branes
wrapped on a Lagrangian submanifold $\S^3$ and another $M$
 non-compact A-branes wrapped on a different Lagrangian submanifold $L=\R^2\times
\S^1$, where these two types of A-branes meet on $\S^1\subset \S^3$. As alluded in
the previous section, after the
geometric transition the  geometry becomes the
resolved conifold, $\CO(-1)\oplus \CO(-1) \rightarrow\P^1$. While
A-branes on $\S^3$ are replaced by the RR-flux on $\P^1$
in this large $N$ transition, $M$ non-compact A-branes wrapped
on $L$ is argued to remain as the
Lagrangian A-branes on the resolved conifold $\widetilde{\cal X}$.

In this case the open string field theory on ${\cal X}$ is reduced to Chern-Simons
theories on two three dimensional Lagrangian submanifold. If we denote the gauge
fields on $\S^3$ and $\R^2\times \S^1$ as $A$ and
$A'$, respectively, then, the effective theory is
described by the action
\beq  S_{CS}[A\,;\, \S^3] + S_{CS}[A'\, ; \, \R^2\times \S^1] +
S_{int}[\phi\,, A\,, A'\,;\S^1]\,, \eeq
where $S_{CS}$ denotes the standard Chern-Simons action. The surviving
degree of freedom of open topological strings for $S_{int}[\S^1]$ is a
complex scalar field
on the circle which is bi-fundamental in the gauge group $U(N)\times
U(M)$. Since $S_{int}[\S^1]$ is quadratic in the complex scalar field,
we can integrate out and obtain the
operator, which was first introduced by  Ooguri and Vafa \cite{Ooguri:1999bv},
\beq Z(U,V) \equiv
\exp\Big[\sum_{n=1}^{\infty}\frac{1}{n}\tr\,U^n~\tr\,V^n\Big]\,,
\label{OV}
\eeq
where ``$\tr$'' denotes the trace for the fundamental
representation and $U$, $V^{-1}$ are the holonomy matrices around $\S^1$
defined by
\beq U\equiv Pe^{\oint_{\S^1}A}\,, \qquad V^{-1} \equiv
Pe^{\oint_{\S^1}A'} \,. \eeq
Using Frobenius character formula, one can get
\beq Z(U,V) = \sum_{\mu} \Tr_{\mu}U\, \Tr_{\mu}V\,, \eeq
where $\Tr_{\mu}$ denotes the trace over an irreducible representation
$\mu$. Since an irreducible representation for $U(N)$ is one-to-one
correspondent with a partition or Young diagram, $\mu$ is specified by
non-decreasing non-negative integers as $\mu=\{\mu_1\ge \mu_2\ge
\cdots\ge \mu_n \ge 0 \}$.

Gauge fields $A'$ on non-compact branes can be treated as
non-dynamical ones or `source' terms from the viewpoint of $U(N)$
gauge theory on $\S^3$.  Therefore the operator (\ref{OV}) may be
regarded as a generating functional for Wilson loop observables in
all representations $\mu$
\beq \langle Z(U,V) \rangle_{\S^3} = \sum_{\mu} \langle
\Tr_{\mu}U\rangle_{\S^3}  \Tr_{\mu}V\,. \label{Unknot1} \eeq
This is another example that the string theory leads to a generating
functional of field theory naturally.
The Wilson loop observable $\langle \Tr_{\mu}U\rangle_{\S^3}$ was
calculated explicitly in \cite{Witten:1988hf} for the unknots as
\beq \langle \Tr_{\mu}U\rangle_{\S^3} =
\frac{S_{\bullet\mu}}{S_{\bullet\bullet}}= s_{\mu}(q^{\frac{N}{2}-i +
  \half})\equiv\dim_q \mu \,,\label{dimension}\eeq
where $S_{\mu\nu}$ is the matrix elements of $S$-transformation of
$SL(2,\Z)$ to get $\S^3$ by gluing two solid torus. The function
$s_\mu(x_i)$ ($i=1,2,\cdots, N$.) is Schur function  introduced in the
appendix with some relevant properties, and $\dim_q \mu $ is the,
so-called, quantum dimension of the representation $\mu$, which
reduces to the ordinary dimension of the representation $\mu$ in the
classical limit $q=e^{g_s}\rightarrow 1$.  The concrete expression of
(\ref{dimension}) can be found in the appendix.

Since the orientation is reversed for anti-branes, the amplitude for
anti-brane insertion is given by
\beq Z^{-1}(U,V) =
\exp\Big[-\sum_{n=1}^{\infty}\frac{1}{n}\tr\,U^n~\tr\,V^n\Big] =
\sum_{\mu}(-1)^{|\mu|}\Tr_{\mu^t}U\Tr_{\mu}V\,, \eeq
which leads to the knot invariants for the unknots
\beq \langle Z^{-1}(U,V) \rangle_{\S^3} = \sum_{\mu}
(-1)^{|\mu|}s_{\mu^t}(q^{\frac{N}{2}-i+\half})  ~\Tr_{\mu}V\,.
\label{Unknot2}\eeq

\section{Non-compact branes in the topological vertex}

Topological  vertex is the main computational tool for the study of
topological A-model on toric Calabi-Yau
space~\cite{Aganagic:2003db}. In this section we describe the
non-compact branes in the context of topological vertex.
Firstly, we describe the topological vertex formalism emphasizing the
role of symmetries. We explain the subtleties in the realization of
these symmetries through the amplitude. We show this in detail for the
case of resolved conifold. Then we consider the insertion of
non-compact branes at the external legs in the toric diagram of the
resolved conifold, which is believed to be the dual of the non-compact
brane configurations in Chern-Simons theory. As will be explained
later, this is deeply related to the mirror B-model where the
amplitude is realized as a wavefunction.

\subsection{The topological vertex}

The toric diagram of the Calabi-Yau space can be decomposed into
the set of vertices which are connected by
lines(propagators)~\cite{Aganagic:2003db}. This is basically
because each vertex represents $\C^3$ and a toric Calabi-Yau space
can be regarded as a union of  $\C^3$ in such a way that $\C^*$
action is well-defined globally. The manifold $\C^3$ may be
regarded as a total space of $\T^2\times \R$ fibration over
$\R^3$, where $\T^2$ fiber space can be taken as the orbit of the
following action
\bea  \alpha &;&  (z_1,z_2,z_3) \longrightarrow (e^{-i\alpha}z_1,z_2,e^{i\alpha}z_3)\,, \nn \\
    \beta &;&  (z_1,z_2,z_3) \longrightarrow (e^{i\beta}z_1, e^{-i\beta}z_2,z_3)\,. \eea
The moment maps for these actions is given by
\beq
r_{\alpha} =|z_3|^2-|z_1|^2\,, \qquad r_{\beta} =|z_1|^2-|z_2|^2\,. \eeq
The trivalent vertex corresponds to the degeneration loci of
$\T^2$ in the $\R^2$ subspace of $\R^3$ base and consists of three
edges or legs denoted as two-dimensional vectors $v_i=(p_i, q_i)$,
satisfying
\begin{equation}
\sum_{i=1}^3 v_i =0~,
\end{equation}
and
\begin{equation}
v_2\wedge v_1 =v_1\wedge v_3 = v_3\wedge v_2 =1~,
\end{equation}
with the wedge product defined by $v_i\wedge v_j = p_i q_j- q_i p_j$.

Now consider an open string amplitude for this vertex with Lagrangian A-branes attached in all
three legs. Then the total open string partition function may be written as
\begin{equation}
Z(V_i) =\sum_{\lambda,\mu,\nu} C_{\lambda\mu\nu}~ \Tr_{\lambda} V_1\,
\Tr_{\mu}V_2\, \Tr_{\nu}V_3\,,
\end{equation}
where $V_i$ is a source associated to the Lagrangian submanifold at
the $i$-th leg and the summation is taken over all possible
irreducible representations. This partition function can be determined
from the link invariants of $U(\infty)$ Chern-Simons theory on $\S^3$
\cite{Aganagic:2003db}. The explicit expression of the topological
vertex amplitude, $C_{\lambda\mu\nu}$, in terms of Schur function is
given in the appendix.

The crucial point is that the topological vertex amplitude which is
obtained from the above computations of open string amplitude,
corresponds to the closed string
amplitude on $\C^3$ with boundaries due to external branes. If we want
to find the A-model closed string partition function on $\C^3$, we
simply take the trivial representation for all three legs.
\vskip5mm
{\it Framing}
\vskip3mm
In the above computations of topological vertex, we used
non-compact Lagrangian branes in $\C^3$.  In order to fully
specify the model, we need to give the boundary conditions on the
fields on these non-compact branes at
infinity~\cite{Aganagic:2000gs}\cite{Aganagic:2001nx}\cite{Marino:2001re}.
This is called a framing in the topological vertex and corresponds
to the framing in the dual Chern-Simons theory. This  boundary
condition can be specified by modifying the geometry and allowing
the $T^2$ fiber to degenerate at additional locations in the base
$\R^3$ so that the Lagrangian branes wrap on compact $\S^3$
cycles. This can be done without affecting the topological A-model
amplitudes. Those additional degeneration loci are specified by
three vectors $f_i$ satisfying
\begin{equation}
f_i\wedge v_i=1~.
\end{equation}
It is clear that the above condition is still satisfied with the
replacement $f_i\rightarrow f_i-n_iv_i$ for any integer $n_i$. A
framing is the choice of the integer $n_i$. We choose the framing,
$(f_1, f_2, f_3)=(v_2, v_3, v_1)$ as the canonical one and denote all
other choices of framing as three numbers $n_i$ relative to this
canonical framing.

The topological vertex in the $(n_1, n_2, n_3)$ framing,
$C^{n_1,n_2,n_3}_{\lambda\mu\nu}$, is related to the topological
vertex in the
canonical one, $C_{\lambda\mu\nu}$, by the relation:
\beq C^{n_1,n_2,n_3}_{\lambda\mu\nu} =
(-1)^{n_1|\lambda|+n_2|\mu|+n_3|\nu|}
q^{\half(n_1\kappa_{\lambda}+n_2\kappa_{\mu}+n_3\kappa_{\nu})}C_{\lambda\mu\nu}\,.
\eeq
%

\vskip5mm
{\it Symmetries}
\vskip3mm
The topological vertex with trivial representation at three legs has
$\widetilde{SL}(2, \Z)$ symmetries inherited from those of $T^2$ fiber
of $\C^3$. Since the wedge product  is invariant under the
$\widetilde{SL}(2, \Z)$ transformation, we can use this
$\widetilde{SL}(2,\Z)$ transformation to change $v_i$ at our
disposal. For example, we can permute $v_i$ cyclically and obtain a
cyclic symmetry $\Z_3$. This $\widetilde{SL}(2, \Z)$ symmetry acts on
the vertex amplitude through the replacement:
\begin{equation}
\label{symmetry}
(f_i, v_i)\longrightarrow ({\tilde g}\cdot f_i,~~ {\tilde g}\cdot
v_i)~,\qquad {\tilde g}\in \widetilde{SL}(2, \Z)
\end{equation}

In what follows, we will consider the $SL(2, \Z)$ transformations acting on $v_i$ only,
\begin{equation}
(f_i, v_i)\longrightarrow (f_i, ~~g\cdot v_i)~,\qquad g\in SL(2, \Z)
\end{equation}
This is in general not a symmetry of the system, thus one should
distinguish it from the $\widetilde{SL}(2,\Z)$ given in
(\ref{symmetry}). One may regard this as a passive transformation.
One can also consider the active transformation acting $f_i$ only,
which moves a non-compact brane at one leg to the one at another
leg, while leaving each leg, $v_i$, invariant. These two
viewpoints is connected by the $\widetilde{SL}(2,\Z)$ symmetry
acting on both $v_i$ and $f_i$ as in (\ref{symmetry}).

The topological vertex amplitude $C_{\lambda\mu\nu}$ is invariant
under the $\Z_3$ subgroup of $SL(2, \Z)$ which takes
\begin{equation}
v_1\rightarrow v_2~, \qquad v_2\rightarrow v_3~, \qquad
v_3\rightarrow v_1~.
\end{equation}
This $\Z_3$ transformation can be realized as $TS^{-1}$ matrices
transforming $v_i=(p_i,q_i)$ as a doublet, where
\begin{equation}
S=\left(\ba{rr} 0 &-1 \\ 1 &0\ea\right),\qquad T=\left(\ba{rr} 1
&1 \\ 0 &1\ea\right)\,.
\end{equation}
This, in turn, becomes an active $ST^{-1}$ transformation of
$SL(2,\Z)$ acting on $f_i$. This results in the cyclic symmetry of
the amplitude $C_{\lambda\mu\nu}$~.
\beq C_{\lambda\mu\nu} = C_{\mu\nu\lambda} = C_{\nu\lambda\mu}\,.  \eeq
The change of framing may also be understood as an active $T$ transformation of $SL(2,\Z)$.

The topological vertex for $\C^3$ has also a $\Z_2$ symmetry from the
exchange of $z_1$ and $z_2$ coordinates of $\C^3$.
\beq C_{\lambda\mu\nu} = q^{\half (\kappa_{\lambda} + \kappa_{\mu} +
\kappa_{\nu})} C_{\lambda^t\nu^t\mu^t}\,.
\label{Z2}\eeq

Note this is not an
$SL(2, \Z)$ transformation.

\vskip5mm
{\it Resolved conifold}
\vskip3mm

The whole closed string amplitude of the toric Calabi-Yau space
containing more than single vertex can be obtained
by gluing appropriately these vertices. The explicit rules for gluing
vertices can be found in \cite{Aganagic:2003db}.
For example, the toric diagram for the
resolved conifold is given by two vertices connected by a line of
length $t$, K\"{a}hler moduli of $\P^1$.

Using the Schur function representation of the topological
vertex given in  the appendix, one
can easily write the vacuum partition function of the closed string for the resolved conifold as
\beq Z^{tv}(g_s,t) =
\sum_{\mu}C_{\bullet\bullet\mu^t}(-e^{-t})^{|\mu|}C_{\mu\bullet\bullet}
= \sum_{\mu}s_{\mu^t}(q^{\rho})s_{\mu}(-e^{-t}q^{\rho})\,, \eeq
where $\mu^t$ denotes the transpose of $\mu$.
It  becomes, through the identity of Schur functions given
in~(\ref{SI2}),
\beq Z^{tv}(g_s,t)= \prod_{i,j=1}^{\infty}(1-e^{-t}q^{-i-j+1})
=\exp\bigg[-\sum_{n=1}^{\infty}\frac{e^{-nt}}{n}\frac{1}{[n]^2}\bigg]\,,
\label{Tvcon}\eeq
where $[n]$ is the so-called $q$-number defined by
\beq [n] \equiv q^{\frac{n}{2}}-q^{-\frac{n}{2}}\,, \qquad  q
\equiv e^{g_s}\,\,. \eeq
Note that $Z^{tv}(g_s,t)$ is identical with
$Z_{\widetilde{\CX}}(g_s,t)$ up to the factor $M(q_c)$. This
non-existence of $M(q_c)$ is a general feature of the topological
vertex calculation and should be taken into account when the
corresponding quantities in the dual theories are compared.

\subsection{Non-compact branes in the topological vertex}

In this section, using topological vertex formalism we compute the
amplitudes of non-compact branes on the resolved conifold at the
various positions.

\vskip5mm
{\it Branes at an external leg}
\vskip3mm

\begin{figure}\center
\includegraphics[width=5cm,height=5cm]{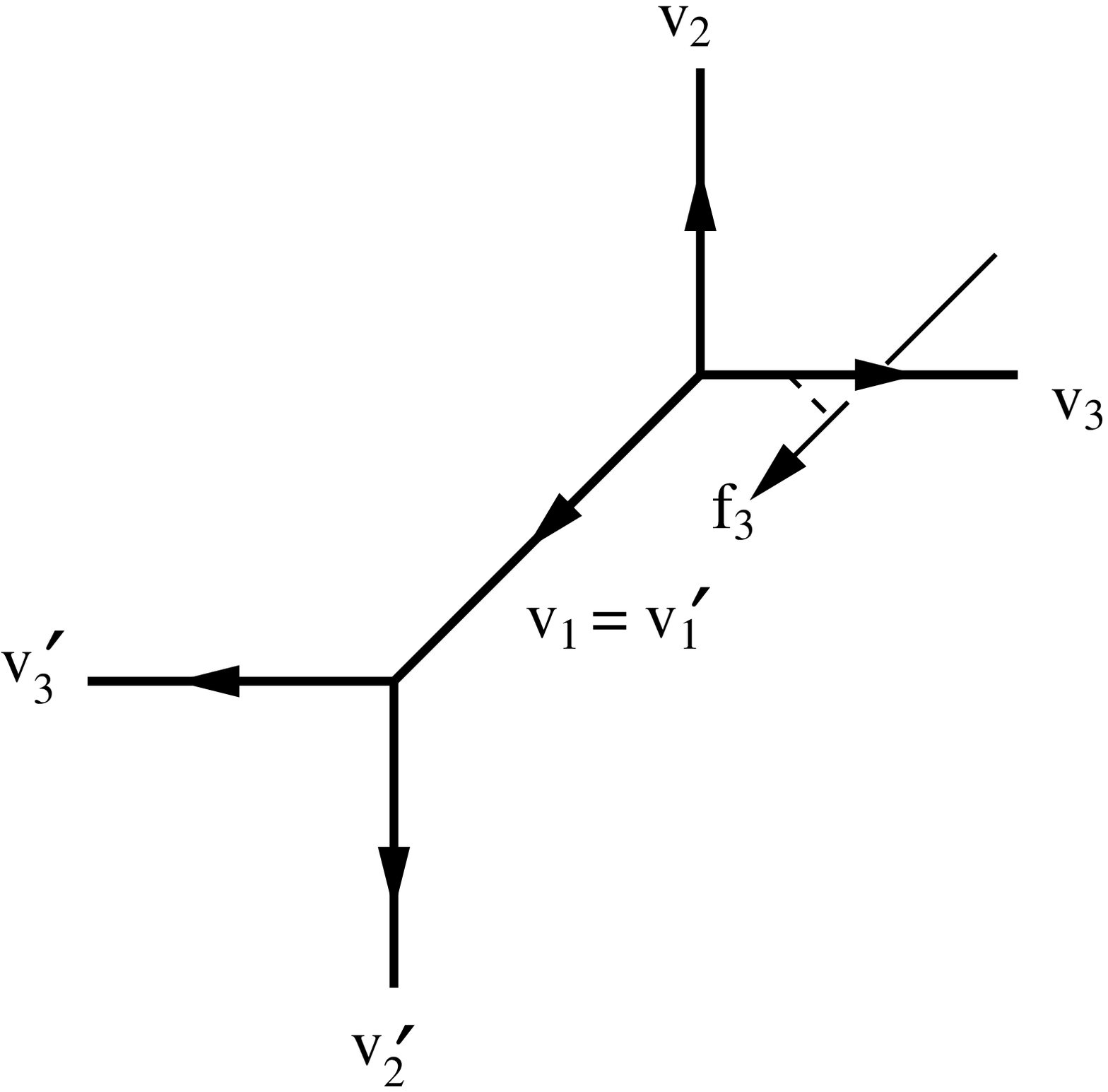} \hskip2cm
\includegraphics[width=5cm,height=5cm]{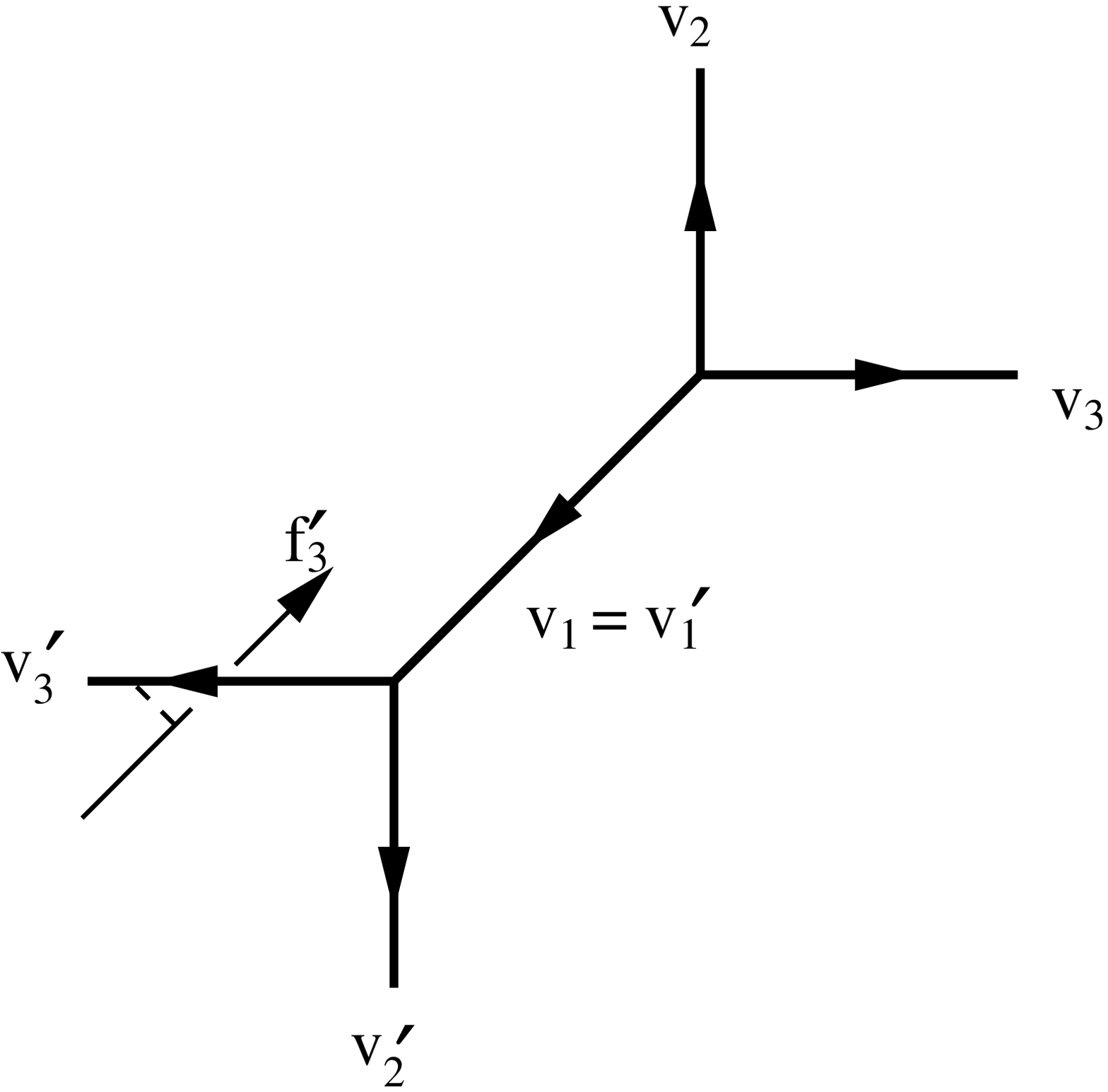}
\caption{Canonically framed A-branes, $f_3$ and $f'_3$, on the external legs $v_3$ and $v'_3$}
\label{f3}
\end{figure}

The amplitude of non-compact branes at an external, non-compact, leg
can be obtained by the replacement
$$C_{\nu\bullet\bullet} \longrightarrow
\sum_{\mu}(e^{-r})^{|\mu|}(-1)^{p|\mu|}q^{\half
  p\kappa_{\mu}}C_{\nu\bullet\mu}\Tr_{\mu}V~,$$
where $p$ denotes a framing.
Therefore the  normalized amplitude for `$f_3$' A-branes with the framing $p$ at the
position `r' on the external leg $v_3$ in the toric diagram for
the resolved conifold, shown in Fig.\ref{f3}, can be calculated in the
formalism of  topological vertex
as
\bea Z^{f_3}_{A}(V,r,p) &=& \frac{1}{Z^{tv}(g_s,t)}\sum_{\mu}\bigg[
\sum_{\nu}(e^{-r})^{|\mu|}(-1)^{p|\mu|}q^{\half p\kappa_{\mu}}
C_{\bullet\bullet\nu^t}(-e^{-t})^{|\nu|}C_{\nu\bullet\mu}\bigg]\Tr_{\mu}V
\nn\\
&=&\frac{1}{Z^{tv}}\sum_{\mu}\bigg[e^{-r|\mu|}(-1)^{p|\mu|}q^{\half
(p+1)\kappa_{\mu}}  \sum_{\nu,\lambda} s_{\nu^t}(-e^{-t}\,
q^{\rho})s_{\mu^t/\lambda}(q^{\rho})
s_{\nu/\lambda}(q^{\rho}) \bigg]\Tr_{\mu}V\,, \eea
where we used Eq.s~(\ref{Hopf}) for
$C_{\nu\bullet\mu}=C_{\bullet\mu\nu}$ and $Z^{tv}$ denotes the vacuum
partition function of A-model topological strings on the resolved
conifold given in~(\ref{Tvcon}).
Using the identities for Schur functions given in Eq.~(\ref{SI2}) and Eq.~(\ref{App}), we obtain
\bea Z^{f_3}_{A}(V, r, p)&=&
\sum_{\mu}\bigg[e^{-r|\mu|}(-1)^{p|\mu|}q^{(p+1)\frac{\kappa_{\mu}}{2}}
\sum_{\lambda}s_{\mu^t/\lambda}(q^{\rho})s_{\lambda^t}(-e^{-t}\,q^{\rho})\bigg]
\Tr_{\mu}V  \nn \\
&=&\sum_{\mu}(-1)^{|\mu|}\bigg[e^{-r|\mu^t|}
(-1)^{(p+1)|\mu^t|}q^{-(p+1)\frac{\kappa_{\mu^t}}{2}}
q^{-\frac{N}{2}|\mu^t|}s_{\mu^t}(q^{\frac{N}{2}-i+\half})\bigg]\Tr_{\mu}V\,,
\eea
where we used $\kappa_{\mu^t}=-\kappa_{\mu}$, $|\mu|=|\mu^t|$, and
$t=g_sN$.
Note that the amplitude of non-compact branes on the external leg $v'_3$
\beq Z^{f'_3}_{A}(V,r,p) = \frac{1}{Z^{tv}}\sum_{\mu}\bigg[
\sum_{\nu}(e^{-r})^{|\mu|}(-1)^{p|\mu|}q^{\half p\kappa_{\mu}}
C_{\bullet\mu\nu^t}(-e^{-t})^{|\nu|}C_{\nu\bullet\bullet}\bigg]\Tr_{\mu}V\,, \eeq
is completely identical with the above $Z^{f_3}$ because of the
symmetry of the topological vertex. Henceforth, it is enough to
consider only one case.

Anti-branes can be obtained by the orientation reversal of branes,
which can be achieved by $v_i \rightarrow -v_i$.
Therefore the amplitude of anti-branes at the position `$r$' on
the external leg $v_3$ is given by
\bea Z^{f_3}_{\bar{A}}(V,
r,p)&=&\frac{1}{Z^{tv}}\sum_{\mu}\bigg[e^{-r|\mu|}
\sum_{\nu}(-1)^{p|\mu^t|}q^{\half p\kappa_{\mu^t}}(-1)^{|\mu|}
C_{\bullet\bullet\nu^t}(-e^{-t})^{|\nu|}C_{\nu\bullet\mu^t}
\bigg]\Tr_{\mu}V \nn \\
&=& \sum_{\mu}\bigg[e^{-r|\mu|}
(-1)^{(p+1)|\mu|}q^{-(p+1)\frac{\kappa_{\mu}}{2}}q^{-\frac{N}{2}|\mu|}
s_{\mu}(q^{\frac{N}{2}-i+\half})\bigg]\Tr_{\mu}V\,. \eea
In contrary to $\C^3$ case, in which the amplitudes for branes and
anti-branes at the same leg are related by the framing change, the
amplitudes for branes and anti-branes are completely different in the
resolved confold case.

\begin{figure}\center
\includegraphics[width=5cm,height=5cm]{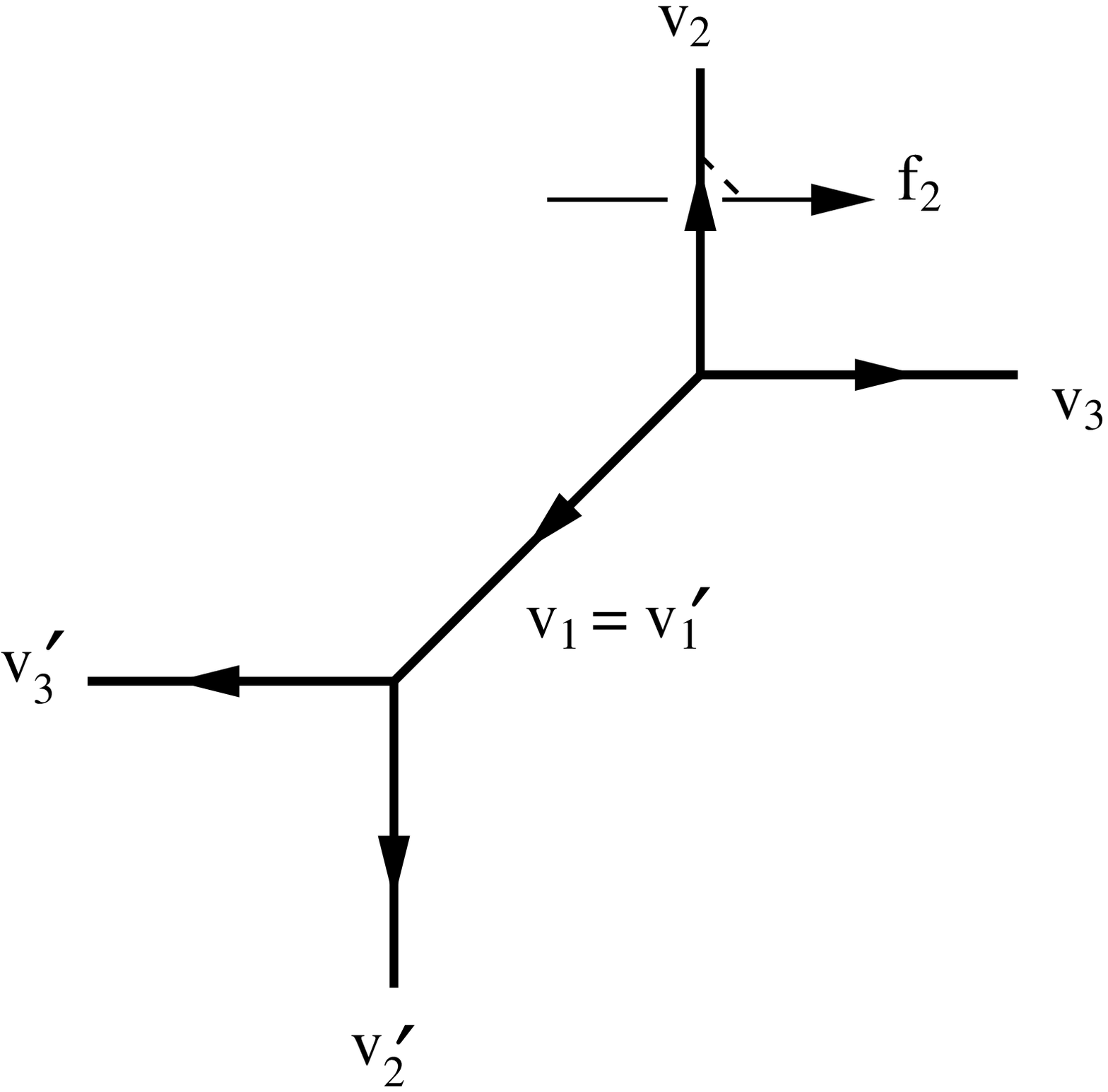} \hskip2cm
\includegraphics[width=5cm,height=5cm]{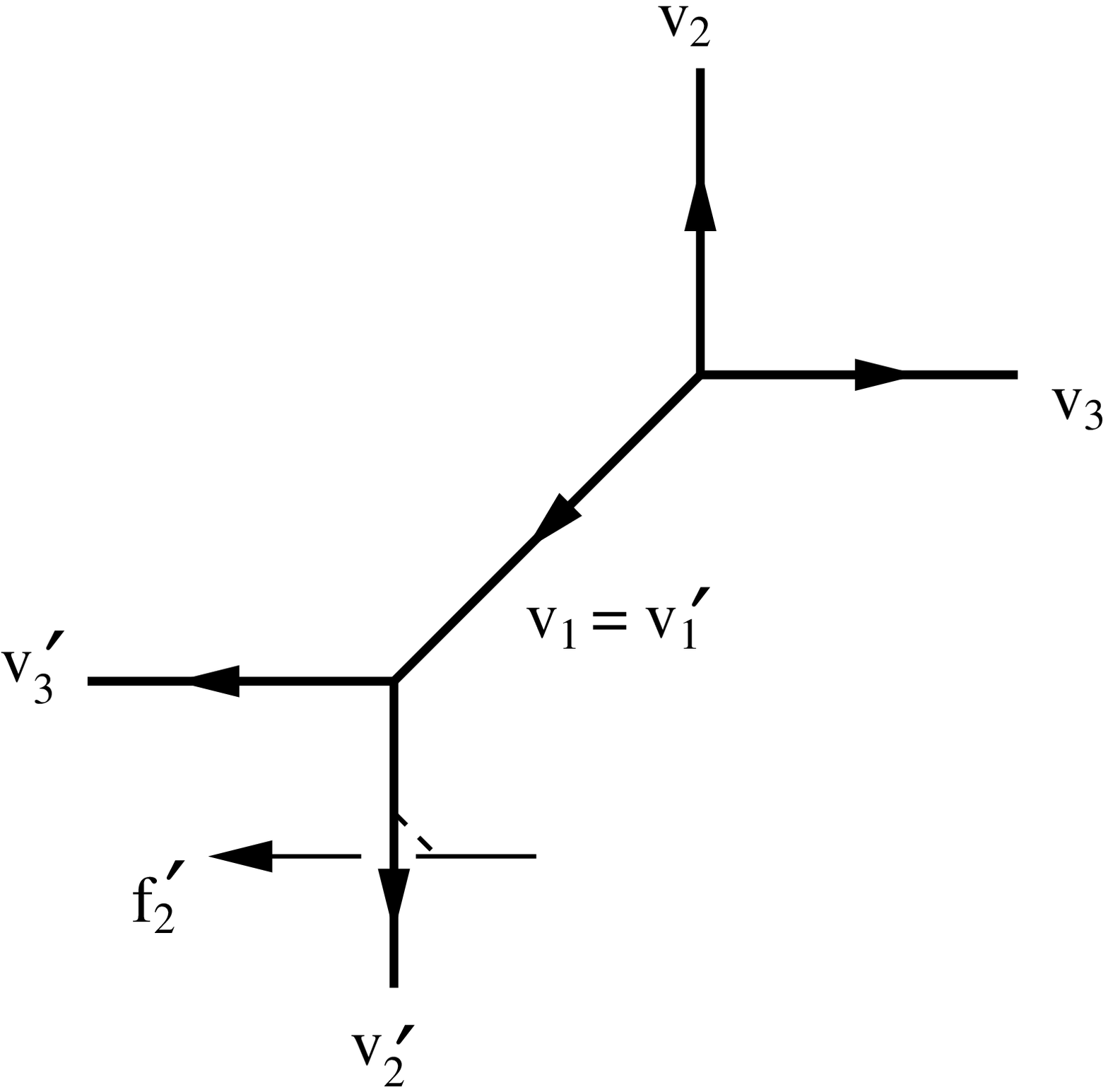}
\caption{Canonically framed A-branes, $f_2$ and $f'_2$, on the external legs $v_2$ and $v'_2$}
\label{f2}
\end{figure}

On the other hand  the amplitudes for the (anti) A-branes at the
external leg $v_2$ or $v'_2$ shown in Fig. \ref{f2} are computed as
\bea Z^{f_2}_{A}(V, r,p) &=&\frac{1}{Z^{tv}}\sum_{\mu}\bigg[
\sum_{\nu}(e^{-r})^{|\mu|}(-1)^{p|\mu|}q^{\half p\kappa_{\mu}}
C_{\bullet\bullet\nu^t}(-e^{-t})^{|\nu|}C_{\nu\mu\bullet}\bigg]\Tr_{\mu}V
\nn\\
&=& \sum_{\mu}\bigg[e^{-r|\mu|} (-1)^{p|\mu|}q^{\half
p\kappa_{\mu}}q^{-\frac{N}{2}|\mu|}
s_{\mu}(q^{\frac{N}{2}-i+\half})\bigg]\Tr_{\mu}V\,,
\label{Amf2}\eea
and
\beq Z^{f_2}_{\bar{A}}(V, r,p) =
\sum_{\mu}(-1)^{|\mu|}\bigg[e^{-r|\mu^t|}
(-1)^{-p|\mu^t|}q^{p\frac{\kappa_{\mu^t}}{2}}q^{-\frac{N}{2}|\mu^t|}
s_{\mu^t}(q^{\frac{N}{2}-i+\half})\bigg]\Tr_{\mu}V\,,
\label{Amf2a}\eeq
respectively.
%
%
Note that there are relations between amplitudes of (anti) A-branes at
the external legs $v_2$ and $v_3$ such that
\beq Z^{f_2}_{\bar{A}}(V,r,p) = Z^{f_3}_{A}(V,r,-p-1)\,, \qquad
Z^{f_2}_{A}(V,r,p) = Z^{f_3}_{\bar{A}}(V,r,-p-1)\,.
\label{AmpZ2}\eeq
In general we don't expect any relation between branes and
anti-branes. However, for the resolved conifold, there is a $\Z_2$
symmetry on the geometry exchanging two external legs, and this
symmetry is realized as the above relations between amplitudes of
branes and anti-branes. It is the same symmetry which exists in $\C^3$
case, alluded earlier in (\ref{Z2}). As we will see, this aspect will be realized
in the mirror B-model setup in the next section, through the
wavefunction behavior of the amplitudes under $\Z_2$.

\vskip5mm
{\it Special cases of the branes at an external leg}
\vskip3mm

For a single (anti)-brane, the holonomy for `source' $V$ is given by
one-dimensional matrix as $V= e^{-i\theta}$, since the gauge group is
$U(1)$. Then, by combining with this Wilson line, the position modulus
$r$ becomes complexified one, $x=r+i\theta$. Furthermore,
representations for $U(1)$ correspond to single row Young diagrams. As
a result, the amplitudes for a single brane and anti-brane
become
\bea Z^{f_3}_{A}(x,\,p)&=& Z^{f_2}_{\bar{A}}(x,-p-1) =
\sum_{n=0}^{N}{N\brack n}q^{-\frac{n}{2}(N-(p+1)(n-1))}
(-1)^{pn}e^{-xn}\,,
\\
Z^{f_3}_{\bar{A}}(x,\,p) &=& Z^{f_2}_{A}(x,-p-1) =
\sum_{n=0}^{\infty}{N+n-1\brack n}
q^{-\frac{n}{2}(N+(p+1)(n-1))}(-1)^{(p+1)n}e^{-xn}\,,
\label{TVAnti}\eea
where ${N+n-1\brack n}$ and ${N \brack n}$ denote quantum dimensions
for a single row and a single column representation. These are
explicitly given by
\bea {N\brack n} &\equiv& \frac{[N]!}{~ [n]!~ [N-n]!} =
q^{\frac{N}{2}n}~ h_{n}(q^{-\half}, q^{-\frac{3}{2}}, \cdots,
q^{-N+\half})\,, \nn \\
{N+n-1\brack n} &\equiv& \frac{[N+n-1]!}{~ [n]!~ [N-1]!} =
q^{\frac{N}{2}n}~ e_{n}(q^{-\half}, q^{-\frac{3}{2}}, \cdots,
q^{-N+\half})\,, \nn \eea
where $h_n$ and $e_n$ are completely symmetric functions and
elementary symmetric functions, respectively, related to Schur functions as
(\ref{SchurHE}) and
\[ [n]! \equiv [n][n-1]\cdots[1]\,, \quad [0]\equiv 1\,.  \]
Note that
\beq \prod_{n=1}^{N}(1-q^{-(n+N_1)})^{-1} =
\sum_{n=0}^{\infty}q^{-n(N_1+\frac{1}{2})}h_{n}(q^{-\frac{1}{2}},q^{-\frac{3}{2}},
\cdots, q^{-N+\frac{1}{2}})\,, \eeq
and
\beq
\prod_{n=1}^{N}(1-q^{-(n+N_1)})=\sum_{n=0}^{N}
q^{-n(N_1+\half)}(-1)^ne_n(q^{-\half},q^{-\frac{3}{2}},\cdots,
q^{-N+\half})\,. \eeq

By taking $N\rightarrow \infty$ limit,  we can reproduce the
(anti-)brane amplitudes on $\C^3$. Note that in the $\C^3$ case
brane amplitudes at framing $p$ are identical with anti-brane
amplitudes at framing $-p-1$ in the same leg due to the combination of
$\Z_2$ and $\Z_3$ symmetry of $\C^3$.

In a particular framing,  these single (anti-)brane amplitudes can
be represented as product forms as
\bea
\label{amp}
Z^{f_3}_{A}(x,\,p=-1)&=& Z^{f_2}_{\bar{A}}(x,p=0)=\prod_{n=1}^{N}(1-e^{-x}q^{-n+\half})\,, \nn
\\
Z^{f_3}_{\bar{A}}(r,\,p=-1) &=& Z^{f_2}_{A}(x,p=0)=\prod_{n=1}^{N}(1-e^{-x}q^{-n+\half})^{-1}\,.
\eea
These forms are relevant for the comparison with the results from
melting crystal picture as will be shown in the next section.

\vskip5mm
{\it Branes at an internal leg}
\vskip3mm

\begin{figure}\center
\includegraphics[width=5cm,height=5cm]{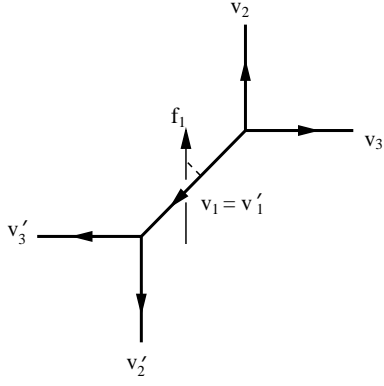}
\caption{Canonically framed A-branes, $f_1$  on the internal leg $v_1$}
\label{f1}
\end{figure}

Now let us consider non-compact branes, $f_1$, on the internal compact
leg, $v_1=v'_1$, of the toric diagram for the resolved conifold in
Fig. \ref{f1}.
In this case, we consider $M$ stacks of a single (anti-)brane with zero
framing, to compare with the results in melting crystal picture.
The amplitude of  $M$ stacks of a single brane on the compact leg at the position $r_a$, $a=1,\cdots, M$ with zero
framing is
\bea Z^{f_1}_{A}(r_a)
&=&\frac{1}{Z^{tv}}\sum_{\mu,\alpha_a,\beta_a}C_{\bullet\bullet\mu\otimes^{M}_{a=1}
\alpha_a}(-e^{-t})^{|\mu|}e^{-\sum_{a}(r_a|\alpha_a|+(t-r_a)|\beta_a|)}C_{\mu^t\otimes^{M}_{a=1}
\beta_a\bullet\bullet} \prod_{a=1}^M\Tr_{\alpha_a}V_a\Tr_{\beta_a}V^{-1}_a \nn \\
&=& \prod_{a=1}^M
\sum_{\alpha_a,\beta_a}s_{\alpha_a}(e^{-r_a}q^{\rho})
s_{\beta_a}(e^{-(t-r_a)}q^{\rho})~ \Tr_{\alpha_a}V_a
\Tr_{\beta_a}V^{-1}_a\,. \eea
Again we can complexify the position modulus $x=r+i\theta$ by
including the Wilson line $V=e^{-i\theta}$ and obtain the amplitude of a single brane
as
\beq \label{v1}Z^{f_1}_{A}(x) = \sum_{m,n=0}^{\infty} e^{-mx}h_m(q^{\rho})~
e^{-n(t-x)}h_n(q^{\rho})=\prod_{m=1}^{\infty}(1-e^{-x}q^{-m+\half})^{-1}\prod_{n=1}^{\infty}
(1-e^{-(t-x)}q^{-n+\half})^{-1} \,. \eeq
%
Similarly, the amplitude of $M$ stacks of a single anti-brane on the compact leg with zero framing at the
position $r_a$ is
\bea Z^{f_1}_{\bar{A}}(r_a) &=& \frac{1}{Z^{tv}}
\sum_{\mu,\alpha_a,\beta_a}C_{\bullet\bullet\mu\otimes
\alpha^t}(-e^{-t})^{|\mu|}C_{\mu^t\otimes
\beta_a^t\bullet\bullet} \prod_{a=1}^M (-1)^{|\alpha_a|+|\beta_a|}e^{-r_a|\alpha_a|-(t-r_a)|\beta_a|}\Tr_{\alpha_a}V\Tr_{\beta_a}V^{-1} \nn \\
&=& \prod_{a=1}^M
\sum_{\alpha_a,\beta_a}s_{\alpha_a^t}(-e^{-r_a}q^{\rho})
s_{\beta_a^t}(-e^{-(t-r_a)}q^{\rho})~ \Tr_{\alpha_a}V
\Tr_{\beta_a}V^{-1}\,. \eea
After the same complexification of the position modulus, the amplitude of an anti-brane becomes
\beq Z^{f_1}_{\bar{A}}(x) =  \sum_{m,n=0}^{\infty} e^{-mx}e_m(-q^{\rho})~
e^{-n(t-x)}e_n(-q^{\rho})=\prod_{m=1}^{\infty}(1-e^{-x}q^{-m+\half})\prod_{n=1}^{\infty}
(1-e^{-(t-x)}q^{-m+\half}) \,. \eeq

All these branes irrespectively to the internal or external leg
insertions have the natural limit to $\C^3$ case by taking ${\rm
Re}\, t$(or $N$) $\rightarrow \infty $. The resultant topological A-model partition function on
$\C^3$ in the presence of the A-brane at the position ``$r$'' with zero
framing\footnote{This is equivalent with anti A-brane with $p=-1$
framing.} can be written as
\beq Z(g_s,r) \equiv e^{F(g_s,r)} =
\exp\bigg[\sum_{n=1}^{\infty}\frac{1}{n}
\frac{e^{-nr}}{2\sinh(\frac{ng_s}{2})}\bigg] =
\exp\bigg[\sum_{n=1}^{\infty}\frac{1}{n}\frac{e^{-nr}}{[n]}\bigg]=
\prod_{k=1}^{\infty}(1-e^{-r}q^{-n+\frac{1}{2}})^{-1}\,. \eeq

\section{Defects in the melting crystal model}


In this section we consider  the melting crystal
model~\cite{Okounkov:2003sp} which seems to be the another
realization of the topological A-model. In particular we consider
(anti-)defects in the melting crystal model which corresponds to
the non-compact branes in the topological A-model on the resolved
conifold~\cite{Saulina:2004da}. Since these defects were
considered recently in~\cite{Halmagyi:2005vk}, we will be brief
and present main results which clearly show the correspondence
with the topological A-model.

As noted earlier, the vacuum partition function of the topological
A-model closed strings on $\C^3$ is given by eq. (\ref{constant}).
This is identical with the partition function of the melting crystal
of cubic lattice in the positive octant of $R^3$,
\beq Z_{crystal}(q_c)=\sum_{3d~ partition} q^{\#\, boxes}_c~, \eeq
where $q_c=e^{-g_s}$.  This observation led to the conjecture that there exists more general melting
crystal picture for topological A-model strings on toric Calabi-Yau
spaces.

The 3d partition can be regarded as the composition of 2d partitions
satisfying the, so-called, interlacing conditions. Conversely
speaking, 2d partitions satisfying the interlacing condition
appear as slicings of the 3d partition. This is clear from the
equivalence between the partition and Young diagrams. However, this
decomposition of the 3d partition into 2d partitions is not unique.
%
We will adopt the convention in which the diagonal
slicing is given by $y=x$ on the positive octant
$(x,y,z)\in\R_{+}^3$.

\begin{figure}[htbp]
\begin{center}
\includegraphics[width=7cm,height=5cm]{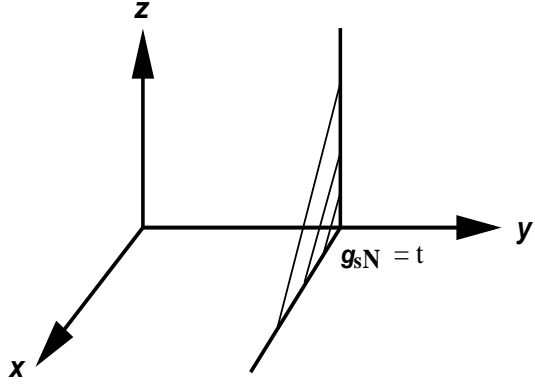}
\caption{melting crystal model for the resolved conifold}
\label{default}
\end{center}
\end{figure}
Two dimensional Young diagram can be represented in terms of a fermionic Fock space. In the transfer matrix formalism, we adopt this and assign a fermionic Fock space to each two dimensional partition. In this formalism we introduce  the operators
$\Gamma_{\pm}$ which satisfy the relations
\beq \Gamma_{+}(z) \Gamma_{-}(z') = \frac{1}{1-z/z'}~
\Gamma_{-}(z')\Gamma_{+}(z)\,, \qquad \Gamma_{+}(z)
\Gamma^{-1}_{-}(z') = \Big(1-\frac{z}{z'}\Big)\,
\Gamma^{-1}_{-}(z')\Gamma_{+}(z)\,. \eeq
These operators
can be realized in terms of a simple harmonic oscillators $\alpha_m$, which are modes of the bosonization of the fermions, as
$\Gamma_{+}(z) = \exp[\sum_{n\ge
1}\frac{\alpha_n}{n}z^n]$ and $\Gamma_{-}(z)  =
\exp[\sum_{n\ge 1}\frac{\alpha_{-n}}{n}z^{-n}]$ with
$[\alpha_m,\alpha_n] = m\delta_{m+n,0}$.

A melting crystal model for the resolved conifold
$\widetilde{\CX}$ is suggested~\cite{Okuda:2004mb} as the lattice in units of $g_s$ in the
positive octant $\R^3_{+}$ with one wall, for
example, at the position $y=g_sN$, see Fig. \ref{default}. The partition function of this
melting crystal model is given by
\beq Z^{crystal}(N,q_c) = \Big\langle 0 \Big| \bigg[
\prod_{n=1}^{\infty}\Gamma_{+}(q_c^{n-1/2})\bigg] \bigg[
\prod_{m=1}^{N}\Gamma_{-}(q_c^{-m+1/2})\bigg] \Big| 0 \Big\rangle
= M(q_c) \prod_{n=1}^{\infty} (1-q_c^{n+N})^n\,. \eeq
One may recall $q_c=q^{-1}$ to see that the partition function  is
identical with $Z_{\widetilde{\CX}}(g_s,t)$ in (\ref{part}) by
performing a resummation and setting $g_sN =t$. In this melting
crystal model, the length from the origin to the wall at the
position $y=g_sN$ corresponds to the K\"ahler moduli $t$ of the
resolved conifold. This suggests that $x$-axis and $y$-axis may
correspond to the external and internal leg, respectively, of
toric diagram for the resolved conifold $\widetilde{\CX}$.

\vskip5mm 
{\it Defects}
\vskip3mm

A defect in the melting crystal model corresponds to a single
non-compact A-brane in the topological A-model. Defect and
anti-defect operators in melting crystals are introduced as
certain fermionic operators
\beq \Psi_{D}(z) \equiv \Gamma^{-1}_{-}(z)\Gamma_{+}(z)\,, \qquad
\Psi_{\bar{D}}(z) \equiv \Gamma_{-}(z)\Gamma^{-1}_{+}(z)\,. \label{defect}\eeq
The partition function with one defect on the $x$-axis at the
position $x=g_s(N_{1}+1/2)$  is given by
inserting $\Psi_{D}(z= q_c^{N_{1}+\half})$ at the appropriate
position as,
\beq Z_{D}(N_1\,;\, N,q_c) = \bigg\langle
0\bigg|\bigg[\prod_{n=N_1+2}^{\infty}\Gamma_{+}(q_c^{n-\half})\bigg]\Psi_{D}(q_c^{N_1+\half})\bigg[\prod_{n=1}^{N_1+1}\Gamma_{+}(q_c^{n-\half})\bigg]\prod_{m=1}^{N}\Gamma_{-}(q_c^{-m+\half})\bigg|
0\bigg\rangle\,. \eeq
Using the commutation relations between $\Gamma$'s we obtain
\beq  Z_{D}(N_1\,;\,N,q_c) =
[\xi(q_c)]^{-1}Z^{crystal}(N,q_c)
\prod_{n=1}^{N}\frac{1}{1-q^{n+N_1}_c}\,. \eeq
where $\xi(q_c) \equiv \prod_{n=1}^{\infty} (1-q^n_c)$. After normalizing by $[\xi(q_c)]^{-1}Z^{crystal}(N,q_c)$,
the amplitude of a defect on the $x$-axis at  $x=g_s(N_{1}+1/2)$
becomes identical with the amplitude, given in eq. (\ref{amp}), of a non-compact brane with $p=0$ at the external
leg $v_2$ (also $v'_2$) with the complexified modulus $x=g_s(N_{1}+1/2)$ in the toric diagram of the resolved conifold, see Fig. \ref{f2}.

It is straightforward, following the argument of the $\C^3$ case in~\cite{Saulina:2004da}, to get the
normalized amplitude for $d$ defects as
\beq  Z_{D}^{norm}(N_i,q_c) = \prod_{1\le i< j \le d}(1-q_c^{N_j-N_i})~
\prod_{i=1}^{d}\prod_{n=1}^{N} \frac{1}{1-q^{n+N_i}_c}\,, \eeq
with $N_i < N_j $ for $ i < j$.

The amplitude for an  anti-defect at $x$-axis can be obtained by inserting $\Psi_{\bar{D}}(q_c^{N_1+1/2})$ instead and is given by
\bea Z_{\overline{D}}(N_1\,;\,N,q_c)
= [\xi(q_c)]Z^{crystal}(N,q_c)\prod_{n=1}^{N} (1-q_c^{n+N_i})\,. \eea
The normalized amplitude of anti-defect is identical with the amplitude of a non-compact anti-brane with $p=0$ at the external
leg $v_2$ (also $v'_2$) in the toric diagram of the resolved conifold.
The normalized amplitude of
multi anti-defects at $x$-axis can be computed in the similar fashion as in the multi defect case and is given by
\beq  Z_{\bar D}^{norm}(N_i,q_c) = \prod_{1\le i< j \le d^t}(1-q_c^{N_j-N_i})~
\prod_{i=1}^{d^t}\prod_{n=1}^{N} (1-q_c^{n+N_i})\,. \eeq

A couple of comments are in order for the amplitude of
(anti-)defects inserted on the $y$-axis. First of all, when we use
the formula for $\Psi_D$ and $\Gamma_{\pm}$, (\ref{defect}), $z$
is determined by the slicing one takes. We use different $z$ for
the defect on the $y$-axis from $z$ for the defect on the
$x$-axis, which are inverse to each other and are due to the
diagonal slicing. Secondly, since the insertion of (anti-)defects
on the $y$-axis corresponds to (anti-)branes in the topological
A-model inserted at $\P^1$, we generally expect the change of
K\"ahler parameter of $\P^1$, and it appears as the shift of the
position of the wall in the melting crystal
model~\cite{Okuda:2004mb}\cite{Halmagyi:2005vk}.

We would like to emphasize that, as will be shown below, the
defect on the $y$-axis corresponds to the anti-brane with $p=0$ at
the compact leg $v_1$ in the toric diagram of the resolved
conifold. This implies that we should interchange the name of
defect and anti-defect operators defined on $y$-axis if we want to
have the correspondence of the naming of brane/anti-brane with
$p=0$ in topological vertex with defect/anti-defect, independent
of the inserted axis. In the case of $\C^3$, this was not clear
since the amplitude of the non-compact brane with $p=-1$ is the
same as the amplitude of the non-compact anti-brane with $p=0$ due
to $\Z_2$ and $\Z_3$ symmetries. By considering more general
space, we could identify the correct defect/anti-defect operators
corresponding to the brane/anti-brane operators.

The amplitude of the defect insertion on the $y$-axis at the position $y = g_s(N_1+1/2)$ can be computed by inserting $\Psi_{D}(z= q_c^{-N_{1}-1/2})$ at the appropriate position and is given by
\bea Z_{D}(N_1 ; N,q_c) =\bigg\langle 0\bigg|
\bigg[\prod_{n=1}^{\infty}\Gamma_{+}(q_c^{n-\half})\bigg]\bigg[\prod_{m=1}^{N_1+1}\Gamma_{-}(q_c^{-m+\half})\bigg]\Psi_{D}(q_c^{-N_1-\half})\prod_{m=N_1+2}^{N+1}\Gamma_{-}(q_c^{-m+\half})\bigg|
0\bigg\rangle~.
\eea
As a result, the normalized amplitude becomes
\bea
Z_{D}^{norm}(N_1\, ;\, N,q_c) =\prod_{m=1}^{\infty}(1-q^{m+N_1}_c)\prod_{n=1}^{\infty}(1-q_c^{n+N-N_1})\,.
\eea
As was mentioned earlier, this corresponds to the topological vertex amplitude (\ref{v1}) of the non-compact anti-branes with $p=0$ inserted at the compact leg $v_1$ in the toric diagram of the resolved conifold.
Similarly, we can compute the normalized amplitude of the anti-defect insertion  on the $y$-axis at the position $y=g_s(N_1+1/2)$  and find
\beq Z_{\bar D}^{norm}(N_1\, ;\, N,q_c) =\prod_{m=1}^{\infty}(1-q_c^{m+N_1})^{-1}
\prod_{n=1}^{\infty}(1-q^{n+N-N_1}_c)^{-1}
    \,. \eeq
This in turn corresponds to the amplitude of the non-compact brane inserted at $v_1$ leg with $p=0$. It is straightforward to generalize those in the above to the amplitude of multi-(anti-)defects insertion at $y$-axis as shown in the case of those at $x$-axis and is omitted.

\section{The B-model on the mirror of the resolved conifold}
In this section we consider the B-model topological string theory on the mirror space of the resolved conifold.  At the beginning we review some salient features of the B-model on some class of the Calabi-Yau spaces. Then we compute the amplitudes of the non-compact brane insertion in the open topological B-model on the mirror space of the deformed conifold in the context of DV matrix model. We regard those as the amplitudes of non-compact branes in the topological B-model on the mirror space of the resolved conifold through the large $N$ duality. We show that they satisfy the Schr\"odinger wave equation confirming the wavefunction nature of the amplitude in the B-model topological string theory.

\subsection{The B-model}

The topological
B-model on the Calabi-Yau space gives the informations on the
complex structure moduli space. The symmetries of the B-model on the
Calabi-Yau space involve holomorphic diffeomorphisms which preserve
the curve equation, like (\ref{mirror}), and the holomorphic 3-form
\begin{eqnarray}\label{3-form}
\Omega =\frac{1}{4\pi^2}\frac{dx\wedge dy \wedge du}{u}~.
\end{eqnarray}
The Calabi-Yau geometry characterized by the curve equation of the form
\begin{equation}
uv= H(x, y)
\end{equation}
can be
regarded as a fibration over the ($x, y$) plane with one
dimensional fibers. The surface satisfying
\begin{equation}
\label{surface}
H(x,y)=0
\end{equation}
 in the base manifold is the locus
where  the fiber degenerates into two components $u=0$ and $v=0$.
It was shown in \cite{Aganagic:2003qj} that this type of CY
geometry is characterized by the algebraic curve (\ref{surface})
and the complex deformations of the CY are captured by the
canonical one-form, $\lambda=ydx$. Therefore if one considers the
deformation of the function $H(x,y)$, while keeping $u, v$ fixed,
the target space theory of B-model, namely Kodaira-Spencer theory,
essentially reduces to the one on the Riemann surface
(\ref{surface}).  The Kodaira- Spencer field  $\phi$ is related to
the one form $\lambda$ by $\lambda=\partial \phi$.  When
restricted to the Riemann surface  (\ref{surface}), they become
the symplectic diffeomorphisms which preserve the equation
(\ref{surface}) and symplectic two-fom $dx\wedge dy$ of the base
manifold. In the quantum Kodaira-Spencer theory, they are realized
as the $\CW$-algebra symmetry.

The Riemann surface satisfying $H(y_i)=0$ in (\ref{LG})
is a genus 0 surface, i.e. a sphere, with four boundaries or
punctures. Near each puncture we can choose a local coordinate $x$ and its conjugate pair $y$ such that $x\rightarrow \infty$ and $y\rightarrow 0$ at the puncture (see, for example, the discussion leading (\ref{LG1})). Since the complex
deformations of the Riemann sphere can arise only at the boundaries,
it is enough to consider the deformation generated by the insertion of
non-compact B-branes near the punctures.  Since $(x,y)$ is a
symplectic pair in the geometry and, furthermore, the action of
non-compact B-branes near the $x$-patch is given by
\begin{equation}
S_B=\frac{1}{g_s}\int y\bar{\partial} x~,
\end{equation}
$y$ naturally plays the role of conjugate momentum of $x$ in the
quantization of branes with the commutation relations
\begin{equation}
[y, x] =g_s~.
\end{equation}
As noted in the above, this
conjugate pair is deeply related to the Kodaira-Spencer field $\phi$
through the one-form $\lambda$ by $\lambda=ydx=\partial \phi$.

Similarly, one can assign an appropriate coordinate and its conjugate momentum near each patch. The non-compact brane near one patch can be moved around and can be connected with the one in another patch by the $SL(2\,,\,\Z)$ coordinate transformations. This $SL(2\,, \Z)$ transformation is the symplectic diffeomorphism
which preserves the symplectic two form $dx\wedge dy$
and the periodicity of the coordinates $x,y$.
There is also freedom in the choice of $(x,y)$, in particular,
coordinate transformation of the form
\begin{eqnarray*}
x\rightarrow x+ ny~, \qquad
y\rightarrow y~,
\end{eqnarray*}
where $n$ is an arbitrary integer. This is also a part of the
symplectic diffeomorphism,  $SL(2\,,\,\Z)$ and is called a framing in
the B-model. As will be clear, it is closely related to the framing in
the mirror topological A-model or topological vertex.

In the case of anti-branes, the action becomes the minus
of the action of branes, $S_{\bar B}=-S_B$, and thus, in the quantization, the conjugate
momentum of $x$ becomes $-y$ which results in the
commutation relation, $[y, x]=-g_s$.

\subsection{Non-compact branes in topological B-model}

Non-compact B-branes inserted near a puncture in the Riemann surface  are described by free chiral
fermions~\cite{Aganagic:2003db} which is related to the Kodaira-Spencer field $\phi$ by $\psi(x)=\exp(\phi(x)/g_s)$. In general, the B-model amplitudes behave as a wavefunction~\cite{Bershadsky:1993cx}. In particular, one-point function of a fermion  which corresponds to the amplitude of single non-compact brane insertion should satisfy the Schr\"{o}dinger equation whose Hamiltonian is given by
the equation of the Riemann surface~\cite{Aganagic:2003qj}:
\beq H(x,~y=g_s\p_x)\langle \psi(x) \rangle = 0\,. \label{Sch}\eeq
The anti-chiral fermion corresponding to an anti-brane can be represented by $\psi^*(x)=\exp(-\phi(x)/g_s)$ and satisfies
\beq H(x,~y=-g_s\p_x)\langle \psi^*(x) \rangle = 0\,. \label{Scha}\eeq

Instead of obtaining the one-point function of a fermion directly from
the closed B-model, we consider the amplitude of the non-compact brane in the context of the large $N$ dual open B-model on $\CY$, which is also the mirror of the deformed conifold. For this purpose it is convenient to use the coordinate transformation alluded earlier in section 2.1 and use the curve equation $ H({\tilde x},{\tilde y}) = 0 $ in eq.(\ref{mirror}) and take the geometric transition.

The resultant open string field theory reduces to DV matrix model with the partition function
\begin{equation}
Z=\frac{1}{{\rm vol} (U(N))}\int d_HU~ \exp\Big(\frac{1}{2g_s} \Tr
U^2\Big)
\end{equation}
where $U$ is a Hermitian matrix and the measure $d_H U$
 is the unitary one~\cite{Aganagic:2002wv}. When expressed in terms of the diagonal
components, the partition function reduces to
\bea
Z=\int \prod_i^N du_i~  \prod_{i<j}\sin^2\Big(\frac{u_i-u_j}{2}\Big)
\exp{\Big[\frac{1}{2g_s}\sum_i u_i^2\Big]}
\eea

In this large $N$ duality, the non-compact B-branes presumably remain
as the non-compact B-branes, meeting the compact one at a point. The
surviving degree of freedom of open string modes connecting these
branes is a bi-fundamental complex scalar whose path integral is given by
\beq \int {\cal D}\bar{\phi}{\cal D}\phi~
\exp\Big[-\bar{\phi}(V\otimes {\bf1}_{N\times N} - {\bf
1}_{M\times M}\otimes U)\phi\Big] = \Big[\det(V\otimes
{\bf1}_{N\times N} - {\bf 1}_{M\times M}\otimes U)\Big]^{-1}\,,
\eeq
where $U$ and $V$  are  matrices from compact and non-compact branes, respectively.

If we consider $M$  non-compact branes at the $M$ different positions
as {$e^{-v_1}, \cdots e^{-v_M} $} on $\P^1$, the matrix $V$ becomes $V=\diag(e^{-v_1}, \cdots e^{-v_M} )$ and  thus we have

\beq
\Big\langle \frac{1}{\det (e^{-v_1}-U)} \frac{1}{\det (e^{-v_2}-U)}\cdots
\frac{1}{\det (e^{-v_M}-U)} \Big\rangle\,
.  \eeq
where the expectation value is taken with respect to $U$. Therefore the amplitude of a non-compact holomorphic brane  is given by
$ \Big\langle \frac{1}{\det (e^{-v}-U)}\Big\rangle $ and using
\beq \langle \Tr_{\mu}U\rangle =
q^{\frac{N}{2}|\mu|}q^{\frac{\kappa_{\mu}}{2}} \dim_q\mu\,, \qquad  q=e^{g_s} \eeq
we obtain
\beq \Big\langle \frac{1}{\det (e^{-v}-U)}\Big\rangle =
\sum_{n=0}^{\infty}{N+n-1\brack
n}q^{\frac{n}{2}(N+n-1)}(e^{v})^{N+n}\,. \eeq

One may consider this as
the amplitude of a non-compact brane in the topological B-model on ${\widetilde \CY}$
through the large $N$ duality
\begin{equation}
\langle \psi({\tilde x}) \rangle=  \Big\langle \frac{1}{\det (e^{-{\tilde x}}-U)}\Big\rangle
\end{equation}

Indeed it satisfies the Schr\"{o}dinger equation (\ref{Sch})
where the Hamiltonian, inherited from the curve equation (\ref{mirror}), is given by
\beq
H({\tilde x}, {\tilde y}=g_s\p_{\tilde x})= q^N-e^{g_s\p_{\tilde x}}-q^{\half}e^{\tilde x}e^{g_s\p_{\tilde x}}+q^{\half}e^{\tilde x}e^{2g_s\p_{\tilde x}}
\,. \label{HamDV1}\eeq

The relevant open string mode connecting the compact brane and the non-compact anti-brane is a bi-fundamental complex fermion. Therefore the path integral of  the amplitude of the insertion of an anti-brane is given by the normalized
expectation value of determinants as
\beq \langle \det (e^{-v}-U)\rangle =
\sum_{n=0}^{N}{N\brack
n}q^{\frac{n}{2}(N-n+1)}(-1)^n(e^{-v})^{N-n}\,. \eeq
The large $N$ dual amplitude $\langle \psi^*({\tilde x}) \rangle = \langle\det (e^{-{\tilde x}}-U)\rangle$ satisfies the Schr\"{o}dinger equation (\ref{Scha}) with the Hamiltonian
\beq H({\tilde x}, {\tilde y}=-g_s\p_{\tilde x})=
q^N-e^{-g_s\p_{\tilde x}}-q^{-\half}e^{\tilde x}e^{-g_s\p_{\tilde
x}}+q^{-\half}e^{\tilde x}e^{-2g_s\p_{\tilde x}} \,.
\label{HamDV}\eeq

There is a different version of the matrix model\footnote{The
unitary matrix model version of the above DV matrix model is also
obtained in~\cite{Okuda:2004mb}.} dual to the closed
B-model~\cite{Tierz:2002jj}. This model is a Hermitian matrix
model with  logarithmic action whose path integral is given by $
\int dM e^{-\frac{1}{2g_s}\Tr(\ln M)^2}\,. $ In this matrix model,
the correlation functions of determinants are given by the Slater
determinant of orthogonal polynomials. In the case at hand, the
relevant orthogonal polynomials are known as Stieltjes-Wigert
ones. In particular, the matrix model vacuum expectation value of
an anti-brane is
\beq \langle \det (e^{-x}-M)\rangle =
(-1)^Nq^{\frac{1}{8}N(N-3)-\frac{1}{4}}\sum_{n=0}^{N}{N\brack
n} q^{-\frac{n}{2}(n+N)}(-q^{-\half}e^{-x})^n\,, \eeq
which is identical to $\langle\det (e^{-v}-U)\rangle$ with identification $x=v-g_s N$ up to an overall constant factor.

\section{Equivalence among various models}

In this section we compare the amplitudes of non-compact branes in
various models. We find the exact correspondence among the
non-compact branes in the  topological string theories.

\vskip5mm
{\it The coordinate patches of Riemann sphere with four punctures}
\vskip3mm

We describe the coordinate patches of Riemann sphere
characterized by the curve equation
\begin{eqnarray}
e^{-Y_1}+e^{-Y_2}+e^{-Y_3}+e^{-Y_4}=0 \nonumber \\
Y_1 +Y_2=Y_3 +Y_4+t
\end{eqnarray}
in the homogeneous coordinates.
 It has four
boundaries and there is an appropriate coordinate near each boundary.

Introduce the coordinates $\{u_i\}$,
\begin{eqnarray}
u_1 =Y_4-Y_3~, \qquad  u_2 =Y_3-Y_1~, \nonumber \\
u_3 =Y_3-Y_4~, \qquad  u_4 =Y_4-Y_2~,
\end{eqnarray}
in which the constraints become
\begin{eqnarray}
u_2 +u_4 +t=0~, \qquad u_1+u_3=0~.
\end{eqnarray}
Then each boundary patch corresponds to the region where
$u_i\rightarrow \infty$ and is described by the local coordinate
$x_i=u_i+\pi i$, which is the natural `flat' coordinate related to
the integrality structure of A-model~\cite{Aganagic:2001nx}. One
can find the canonical conjugate momentum $y_i$ in each coordinate
patch by requiring $y_i\rightarrow 0$ as $x_i\rightarrow \infty$.
The coordinate and conjugate momentum and the curve equation in
each patch are displayed in the table \ref{patch}.

\begin{table}[htdp]
\caption{coordinate patches}
\begin{center}
\begin{tabular}{|c|c|c|}
\hline
coordinate $x$& momentum $y$& curve \\ \hline
 $u_1+\pi i$ &  $u_2+\pi i$ & $-e^{-x}-e^y+e^{-x-y-t}+1=0$\\ \hline
 $u_2 +\pi i$ & $-u_1-u_2+\pi i$ & $-e^{-x}-e^y+e^{-x+y-t}+1=0$\\ \hline
$u_3+\pi i$ & $u_4+\pi i$ &$-e^{-x}-e^y+e^{-x-y-t}+1=0$\\ \hline
$u_4+\pi i$ & $-u_3-u_4+\pi i$ &$-e^{-x}-e^y+e^{-x+y-t}+1=0$\\ \hline
\end{tabular}
\end{center}
\label{patch}
\end{table}%

\vskip5mm
{\it SL(2,Z) transformation in the Riemann sphere}
\vskip3mm

The coordinate and momentum in each patch are related by $SL(2,\Z)$ transformations which preserve the symplectic two form and the periodicity of the coordinates $x_i$ and momenta $y_i$.
 The coordinate transformations from $u_1$-patch to  $u_2$
and from  $u_3$-patch to  $u_4$ are given by $SL(2, \Z)$
transformation $S^{-1}T$, while
the coordinate transformations from
$u_2$-patch to  $u_3$ and from  $u_4$-patch to  $u_1$ are given by
$T^{-1}S^{-1}$ transformation supplemented by constant shift.

\beq
{x_1\choose y_1}\stackrel{S^{-1}T}{\longrightarrow}  {x_2\choose y_2}
\stackrel{T^{-1}S^{-1}}{\longrightarrow}
{x_3\choose y_3}\stackrel{S^{-1}T}{\longrightarrow}
{x_4\choose y_4} \stackrel{T^{-1}S^{-1}}{\longrightarrow} {x_1\choose y_1}\,.
\eeq
%

\begin{figure}[htbp]
\begin{center}
\includegraphics[width=7cm,height=5.5cm]{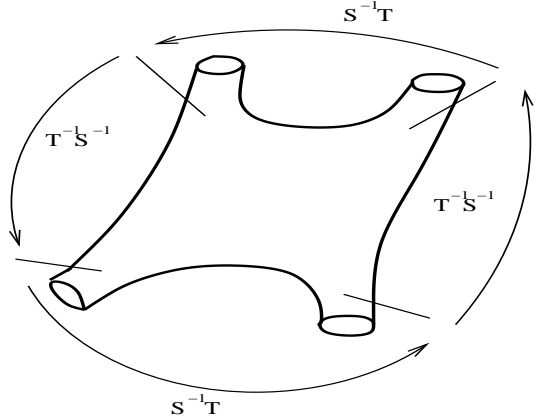}
\caption{$SL(2\,, \Z)$ transformations relating coordinates among
patches in the Riemann surface} \label{bsl2z}
\end{center}
\end{figure}

\vskip5mm
{\it SL(2,Z) transformation in the toric diagram}
\vskip3mm

\begin{figure}[htbp]
\begin{center}
\includegraphics[width=7cm,height=5.5cm]{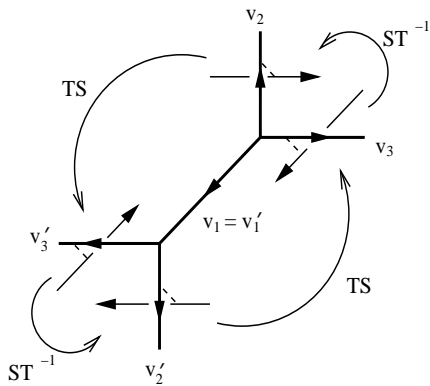}
\caption{Active $SL(2\,, \Z)$ transformations acting on branes at
the external legs in the toric diagram of the resolved conifold,
$\widetilde{\CX}$} \label{sl2z}
\end{center}
\end{figure}

As alluded earlier, the topological vertex also has $SL(2\,, \Z)$
transformations acting on the leg $v_i=(p_i,q_i)$. In the resolved
conifold case, these transformation can be considered acting on
the four external legs $v_{2,3}$ and $v'_{2,3}$ as
\beq
{p_3\choose q_3}\stackrel{TS^{-1}}{\longrightarrow}  {p_2\choose q_2}
\stackrel{S^{-1}T^{-1}}{\longrightarrow}
{p'_3\choose q'_3}\stackrel{TS^{-1}}{\longrightarrow}
{p'_4\choose q'_4} \stackrel{S^{-1}T^{-1}}{\longrightarrow} {p_3\choose q_3}\,,
\eeq
which is exactly in one-to-one correspondence with the
transformations in the Riemann sphere, and can be realized, in the
active view point,  as moving A-branes from one leg to another one
as(see, Fig. \ref{sl2z}.)
\beq
 f_3\stackrel{ST^{-1}}{\longrightarrow}  f_2
\stackrel{TS}{\longrightarrow}
f'_3\stackrel{ST^{-1}}{\longrightarrow} f'_2
\stackrel{TS}{\longrightarrow} f_3\,. \eeq
%

The correspondence between the transformations of A- and B-models
can be easily inferred from the moment map of the $\T^2$ action
for the resolved conifold
\beq r_{\alpha} (\phi)= |\phi_4|^2-|\phi_1|^2\,, \qquad    r_{\beta}= |\phi_1|^2-|\phi_3|^2\,, \eeq
which can also be written as
\beq r_{\alpha} (\phi)= |\phi_2|^2-|\phi_3|^2-{\rm Re}\, t\,,
\qquad    r_{\beta}= |\phi_4|^2-|\phi_2|^2+{\rm Re}\,t \,. \eeq
For example, the coordinate and momentum in the $u_1$-patch  of
the Riemann surface is given by $(x_1, y_1)=(Y_4-Y_3+\pi i,\,
Y_3-Y_1+ \pi i)$ whose real part is nothing but $(r_\alpha
+r_\beta, -r_\beta)$. If the same transformation matrices are
applied on $(r_\alpha +r_\beta, -r_\beta)$ as those on $(x_1,
y_1)$, namely $S^{-1}T$, then naturally the transformation
matrices for $v_3=(p_3, q_3)$ leg become $TS^{-1}$. Therefore one
may regard this exact correspondence between the transformation
rules of both sides as the consequence of the mirror symmetry
between those two Calabi-Yau spaces ${\widetilde \CX}$ and
${\widetilde \CY}$.

\vskip5mm
{\it The wavefunction of a non-compact brane and mirror symmetry}
\vskip3mm

We obtained the A-model amplitudes of the non-compact brane
inserted at the various external legs of the toric diagram for the
resolved conifold in section 3. Now we verify that these
amplitudes exactly correspond to the B-model amplitudes of the
non-compact brane inserted near the punctures of the Riemann
surface in the mirror manifold of the resolved conifold, thus
confirming the mirror symmetry. This can be achieved by showing
that the A-model amplitudes satisfy the Schr\"odinger equation
whose Hamiltonian is given by the curve equation $H(x,y)=0$ with
the replacements $y=g_s\p_x$ for branes and $y=-g_s\p_x$ for
anti-branes.

Essentially there are only two different realizations of the curve
equation as is clearly seen in the Table  \ref{patch}. One can see
that the brane amplitude at the $v_3$ leg in the toric diagram for
the resolved conifold satisfies
\beq \Big(1-e^{g_s\p_x}-(-1)^pq^{-\half}e^{-x}e^{-pg_s\p_x}
+(-1)^pq^{-N-\half}e^{-x}e^{-(p+1)g_s\p_x}\Big)
Z^{f_3}_{A}(x,p)=0\,. \eeq
This tells us that $Z^{f_3}_{A}(x,p)$ is exactly the one-point
function of the B-model non-compact brane/fermion operator at the
$u_1$ patch, as it satisfies the Schr\"odinger equation with the Hamiltonian 
inherited from the curve equation written in the variables for $u_1$ patch with a framing $p$. One can also check that the anti-brane amplitude  at
the external leg $v_3$ obeys the same Schr\"odinger equation 
\beq \Big(1-e^{-g_s\p_x}-(-1)^pq^{\half}e^{-x}e^{pg_s\p_x}+(-1)^p
q^{-N+\half}e^{-x}e^{(p+1)g_s\p_x}\Big)
Z^{f_3}_{\bar{A}}(x,p)=0\,, \eeq
where now we replaced $y\rightarrow -g_s\p_x$ appropriate for anti-branes.
Alternatively, after factoring out differential operators, one can show that it satisfies 
\beq
\Big(1-e^{g_s\p_x}+(-1)^pq^{-\half}e^{-x}e^{(p+1)g_s\p_x}-(-1)^p
q^{-N-\half}e^{-x}e^{(p+2)g_s\p_x}\Big)
Z^{f_3}_{\bar{A}}(x,p)=0\,. \label{AF3}\eeq
The $\Z_2$ symmetry of the resolved conifold, which is realized on the
brane amplitude as eq.~(\ref{AmpZ2}) implies that similar
differential equations hold for the amplitude of
non-compact A-branes at another legs in the toric diagram for the
resolved conifold. For example, it can be deduced from eq.~(\ref{AF3}) that the amplitude of non-compact A-branes at the external leg $v_2$ satisfies
\beq \Big(1-e^{g_s\p_x}-(-1)^pq^{-\half}e^{-x}e^{-pg_s\p_x}+(-1)^p
q^{-N-\half}e^{-x}e^{(1-p)g_s\p_x}\Big) Z^{f_2}_{A}(x,p)=0\,, \eeq
which is nothing but the Schr\"odinger equation written in the variables appropriate for $u_2$ patch. All other cases can be
deduced easily, too. 

These tell us that the A-model amplitudes are  the same as the B-model amplitudes supporting the mirror symmetry. We will see this explicitly in below. Note that the $\Z_2$ symmetry is the isometry of the resolved conifold and its mirror CY space. It is realized  as  the relation among the (anti-)brane
amplitudes in the A-model on the resolved conifold and this should also be the case for the amplitudes in the mirror B-model. Furthermore, it is well known that the amplitudes in the B-model behave like wavefunctions. Our computation  clearly shows that  the A-model amplitudes have the
properties of the wavefunction, just like the amplitudes in the
B-model.

\begin{figure}[htbp]
\begin{center}
\hskip2.5cm
\includegraphics[width=6cm,height=5cm]{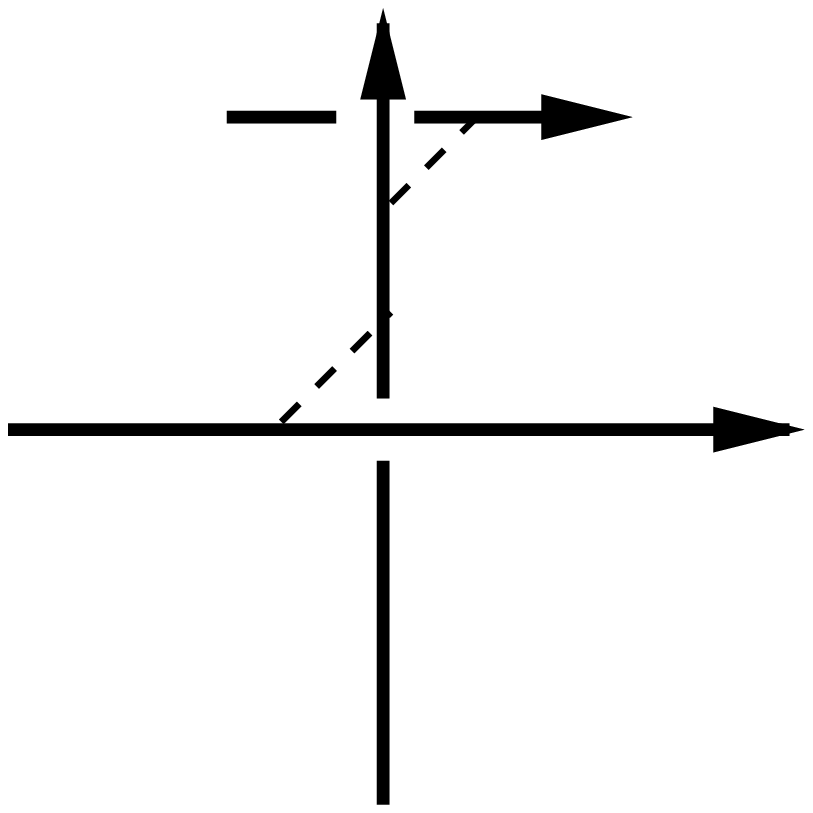}
\includegraphics[width=6cm,height=5cm]{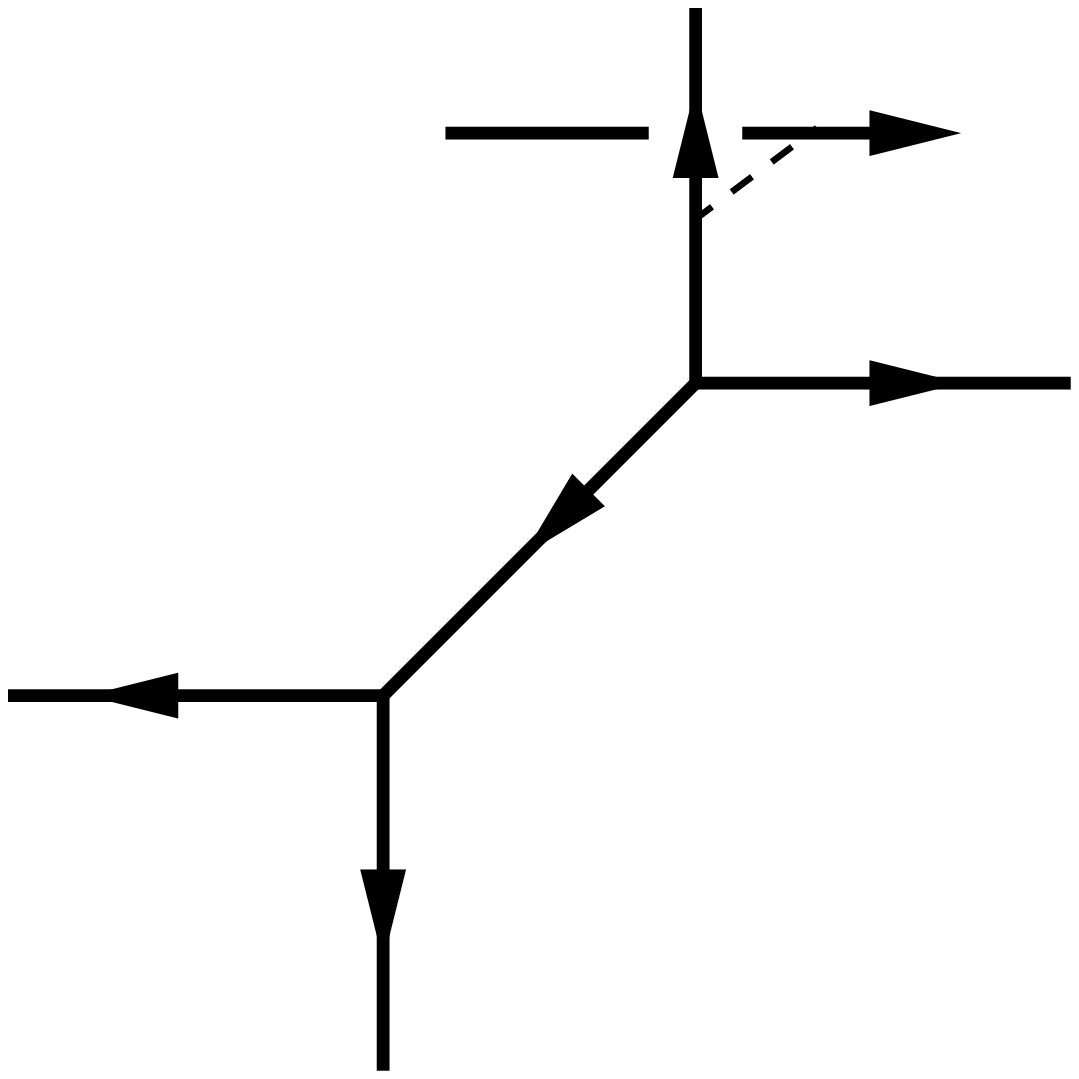}
\caption{L.H.S: A-branes on the deformed conifold.  R.H.S:
A-branes on the resolved conifold after the geometric transition}
\label{trans1}
\end{center}
\end{figure}

\vskip5mm {\it Chern-Simons theory vs. topological vertex vs. DV
matrix model} \vskip3mm

The A-model topological open string field theory on the deformed
conifold becomes the $U(N)$ Chern-Simons theory on $\S^3$. On the
other hand, it is believed that the A-model topological  string
theory on the toric Calabi-Yau space can be described by the
topological vertex. There is a large $N$ duality between the
A-model topological open string theory on Lagrangian branes
wrapped on $\S^3$ in the deformed conifold and the A-model
topological  string theory on the resolved conifold which is
related to the former via geometric transition.

We can see from eq.s~(\ref{Amf2}) and (\ref{Amf2a}) that the knot invariants of the unknot in Chern-Simons theory given in~eq.s (\ref{Unknot1}) and (\ref{Unknot2}) are
identical with the amplitude of non-compact (anti-)branes inserted
at the external leg $v_2$ with $p=0$ and $r= - g_sN/2$:
\[ \langle Z(U,V)\rangle_{\S^3} = Z^{f_2}_{A} \Big(V,~ r =
-\half g_sN, p=0\Big)\,, \qquad  \langle Z^{-1}(U,V)\rangle_{\S^3}
= Z^{f_2}_{\bar{A}} \Big(V,~ r = -\half g_sN, p=0\Big)\,. \]
This correspondence between the Chern-Simons knot invariants and the
amplitudes at the leg $v_2$ is pictorially manifest from Fig.
\ref{trans1}. We can also find a complete agreement for the knot
invariants at a different framing in Chern-Simons theory by taking
a suitable framing and an appropriate assignment of the position
modulus $r$ in topological vertex.

In the previous sections we computed the amplitudes of a
non-compact brane in the A-model and the B-model, independently.
In the B-model we obtained the amplitude from the large $N$ dual
DV matrix model. After an appropriate normalization, the B-model
amplitude of a non-compact brane inserted at $\P^1$ is given by
the determinant expression. Again we see that there is a similar correspondence between the amplitudes in the DV matrix model and those in
topological vertex. As alluded earlier, it  is clear from the Schr\"odinger equation
which they satisfy. More precisely, if we use the `normalized'
expression in DV matrix model, we can see that\footnote{There was
observation~\cite{Okuyama:2006eb} that there is discrepancy
between the results from melting crystal model, which agrees with
topological vertex, and those from DV matrix model. This is just
framing difference between branes  on the $u_2$-patch coming from
DV matrix model and the one from the melting crystal model.}
\bea \Big\langle \frac{\det e^w}{\det( e^w - U)} \Big\rangle &=&
Z^{f_2}_{A}(x=w-g_sN+\pi i,~ p=1)\,, \nn \\
\Big\langle \frac{\det( e^w - U)}{\det e^w} \Big\rangle &=&
Z^{f_2}_{\bar{A}}(x=w-g_sN+\pi i,~ p=1)\,. \eea
As was shown in~(\ref{HamDV1}) and (\ref{HamDV}), the unnormalized amplitude satisfies
the Schr\"odinger equation coming from  the curve equation $H(\tilde{x},\tilde{y})=0$ in~(\ref{mirror}),
which is another representation of the curve in the same $u_2$-patch. But this curve equation is related by a suitable
framing change and rescaling from our choice of coordinates for
the $u_2$-patch, which explains the rescaling in the
B-model amplitude for the complete agreement with the topological
vertex result at the $v_2$ leg(or $u_2$-patch in the corresponding
Riemann surface) with framing change. 
After this identification, we can obtain all the other amplitudes of non-compact branes in different patch by suitable $SL(2,\, \Z)$ transformations.

\section*{Acknowledgments}
The work of S.H. was supported in part by 
grant No. R01-2004-000-10651-0 from the Basic Research Program of the
Korea Science and Engineering Foundation (KOSEF) and by the Science Research Center Program of 
the Korea Science and Engineering Foundation through 
the Center for Quantum Spacetime(CQUeST) of 
Sogang University with grant number R11 - 2005 - 021.
 The work of S.-H.~Y was supported by
the Korea Research Foundation Grant KRF-2005-070-C00030.

\appendix
\section{Schur functions}
In this appendix we present various properties and formulae of Schur functions, which are used in the computation of the topological vertex. Although most of them are just the summary of well-known results~\cite{Mac}, one useful identity is presented with a proof, which does not seem to be written down explicitly in the literatures.

Schur
functions for a partition $\mu = (\mu_1,\mu_2,\cdots,\mu_d)$ may be defined as
\beq s_{\mu}(x_i) \equiv s_{\mu}(x_1,\cdots,x_N) = \frac{\det
x_j^{\mu_i+N-i}}{\det x_j^{N-i}}\,,  \label{SchurDef}\eeq
which can also be represented in terms of elementary or completely
symmetric functions as
\beq s_{\mu}(x_i) = \det (e_{\mu^t_i-i+j})= \det
(h_{\mu_i-i+j})\,.  \label{SchurHE}\eeq
These symmetric functions $h_n$ and $e_n$ can be defined as
\bea \prod_{i=1}^{N}(1-x_i\, t )^{-1} &=&
\sum_{n=0}^{\infty}t^{n}h_{n}(x_i)\,, \nn \\
\prod_{i=1}^{N}(1+x_i t) &=& \sum_{n=0}^{N} t^ne_n(x_i)\,. \eea

Note that $s_{\mu}$ is homogeneous of degree $|\mu|$. For the
explicit expression of Schur functions relevant to our cases, it
is convenient to introduce the {\it content} $c(a)$, and the {\it
hook-length} $h(a)$, at the position $a=(i,j)\in \mu$ of a Young diagram
$\mu$, which are defined by
\[
 c(a)\equiv j-i\,, \qquad h(a)\equiv \mu_i+\mu^t_j-i-j+1\,. \]
These $c(a)$ and $h(a)$ satisfy
\beq \sum_{a\in \mu}c(a) = n(\mu^t)-n(\mu)=\half
\sum_{i=1}^{d(\mu)}\mu_i(\mu_i-2i+1) \equiv \half\kappa_{\mu}\,,
\qquad \sum_{a\in \mu}h(a) = \half\kappa_{\mu} +2n(\mu) + |\mu|\,,
\eeq
where $d(\mu)$ denotes the number of rows for the given Young
diagram $\mu$. By specializing $x_i=q^{-i+1/2}$ with the formula
in~\cite{Mac}, one can derive
\beq s_{\mu}(q^{-i+\half}) = q^{-n(\mu)-\half|\mu|}\prod_{a\in
\mu}\frac{1-q^{-N-c(a)}}{1-q^{-h(a)}} =
q^{-\frac{|\mu|}{2}N}\,\prod_{a\in \mu}\frac{q^{\frac{N}{2}+\half
c(a)} - q^{-\frac{N}{2}-\half c(a)}}{q^{\half h(a)} -  q^{-\half
h(a)}}\,, \eeq
where $n(\mu) \equiv \sum_{i=1}^{d(\mu)}(i-1)\mu_i$. Note that
$q^{\frac{N}{2}|\mu|}s_{\mu}(q^{-i+1/2})$ is nothing but the
quantum dimension or the knot invariants for the unknots. The
given form is efficient to get the explicit expression of the
quantum dimension for a Young diagram of the small number of
boxes.

Using another identities given in~\cite{Mac}
\bea \prod_{a\in \mu}(1-t^{h(a)}) &=&
\frac{\prod_{i=1}^{d(\mu)}\prod_{k=1}^{d(\mu)+\mu_i-i}(1-t^k)}{\prod_{1\le
i<j\le d(\mu)}(1-t^{\mu_i-\mu_j+j-i})}\,, \\
\prod_{a\in \mu}(1-t^{N+c(a)}) &=&
\prod_{i=1}^{d(\mu)}\prod_{k=-i+1}^{\mu_i-i}(1-t^{N+k})\,, \nn
\eea
one can get the form of the quantum dimension given
in~\cite{Marino:2004uf}
\bea s_{\mu}(q^{\frac{N}{2}-i+\half}) =\dim_q \mu  &=&
\prod_{1\le i<j\le d(\mu)}[\mu_i-\mu_j+j-i]
\prod_{i=1}^{d(\mu)}\frac{\prod_{k=-i+1}^{\mu_i-i}[k]_{q^{N}}}
{\prod_{k=1}^{d(\mu)+\mu_i-i}[k]} \nn \\
&&\nn \\ &=& \prod_{1\le i<j\le
d(\mu)}\frac{[\mu_i-\mu_j+j-i]}{[j-i]} \prod_{i=1}^{d(\mu)}\frac{
\prod_{k=-i+1}^{\mu_i-i}[k]_{q^{N}}}
{\prod_{k=1}^{\mu_i}[k-i+l(\mu)]} \,, \eea
where $[k]\equiv q^{\frac{k}{2}}-q^{-\frac{k}{2}}$ and $[k]_{q^N}
\equiv q^{\frac{N}{2}+\frac{k}{2}}-q^{-\frac{N}{2}-\frac{k}{2}}$.

Skew Schur functions can be introduced as
\beq s_{\lambda/\mu} = \sum_{\nu}c^{\lambda}_{\mu\nu}s_{\nu}\,,
\eeq
where the `tensor product coefficient' $c^{\lambda}_{\mu\nu}$ are
integers defined by $s_{\mu}s_{\nu} =
\sum_{\lambda}c^{\lambda}_{\mu\nu}s_{\lambda}$. Note that
$s_{\lambda/\mu}$ is a homogeneous function of degree
$|\lambda|-|\mu|$ and  Schur functions can be understood as a
special case of skew Schur functions since
$s_{\lambda/\bullet}=s_{\lambda}$. As in~\cite{Okounkov:2003sp},
$s_{\lambda/\mu}(q^{\mu+\rho})$ represents a  skew Schur function
of infinite number of variables with the specialized
$x_i=q^{\mu_i-i+1/2}$ $(i=1,2,\cdots)$, which can be understood as
taking $N\rightarrow \infty$ in the case of the finite number of
$x_i$. In particular, $s_{\mu}(q^{\rho})$ can be written
explicitly as
\beq s_{\mu}(q^{\rho}) = q^{-n(\mu)-\half|\mu|}\prod_{a\in
\mu}\frac{1}{1-q^{-h(a)}}\,, \eeq
which leads to
\beq s_{\mu^t}(q^{\rho}) = q^{-\kappa_{\mu}/2}s_{\mu}(q^{\rho})\,.
\eeq
By analytic continuation one may see
\beq s_{\mu} (q^{-\rho}) = (-1)^{|\mu|}s_{\mu^t}(q^{\rho})\,. \eeq
%
The definition of Schur functions as a ratio of determinants given
in~(\ref{SchurDef}) leads to the following identity for a finite
number of variables $x_i= q^{N-i},~ q^{\mu_i+N-i}$~
$(i=1,2,\cdots, N)$
\beq s_{\mu}(q^{N-i})s_{\nu}(q^{\mu_i+N-i}) =
s_{\mu}(q^{\nu_i+N-i})s_{\nu}(q^{N-i})\,, \eeq
which gives us in the limit of $N\rightarrow\infty$
\beq s_{\mu}(q^{\rho})s_{\nu}(q^{\mu+\rho}) =
s_{\mu}(q^{\nu+\rho})s_{\nu}(q^{\rho})\,. \eeq
Another useful formulae for skew Scur functions~\cite{Mac}:
\bea \sum_{\lambda} s_{\lambda/\mu}(x)s_{\lambda/\nu}(y) &=&
\prod_{i,j\ge 1}\frac{1}{1-x_iy_j}~
\sum_{\sigma}s_{\nu/\sigma}(x)s_{\mu/\sigma}(y)\,, \label{SI1} \\
\sum_{\lambda}s_{\lambda/\mu^t}(x)s_{\lambda^t/\nu}(y) &=&
\prod_{i,j\ge 1}(1+x_iy_j)~
\sum_{\sigma}s_{\nu^t/\sigma}(x)s_{\mu/\sigma^t}(y)\,.
\label{SI2}\eea
Now, we will derive a useful formula for the topological vertex
calculations. \vskip3mm

{\bf\large Proposition}: The following identity holds
\beq s_{\mu}(x) =\sum_{\nu}s_{\mu/\nu}(x,y)s_{\nu^t}(-y)\,. \eeq
{\bf Proof}: Let us start from the following identity
(see,~\cite{Mac})
\[ s_{\lambda}(x,y) = \sum_{\nu}s_{\lambda/\nu}(x)s_{\nu}(y)
 = \sum_{\nu,\mu,\lambda}c^{\lambda}_{\nu\mu}s_{\mu}(x)s_{\nu}(y)\,.
\]
Multiplying $s_{\lambda}(y)$ and summing over $\lambda$, one get
\bea \sum_{\lambda} s_{\lambda}(x,y) s_{\lambda}(y) &=&
\sum_{\nu,\mu,\lambda}c^{\lambda}_{\nu\mu}s_{\mu}(x)s_{\nu}(y)s_{\lambda}(y)
\nn\\ &=& \sum_{\nu}s_{\nu}(y)s_{\nu}(y)
\sum_{\mu}s_{\mu}(x)s_{\mu}(y)\,.  \nn \eea
Using eq.s~(\ref{SI1}) and (\ref{SI2}), one get
\[ \sum_{\nu}s_{\nu}(y)s_{\nu}(y)  =
\Big[\sum_{\nu}s_{\nu}(y)s_{\nu^t}(-y)\Big]^{-1}\,,
\]
which leads to
\bea \sum_{\mu}s_{\mu}(x)s_{\mu}(y) &=& \sum_{\lambda}
s_{\lambda}(x,y) s_{\lambda}(y)
\sum_{\nu}s_{\nu}(y)s_{\nu^t}(-y)\,, \nn\\
&=& \sum_{\mu}\Big[\sum_{\lambda,\nu}c^{\mu}_{\nu\lambda}
s_{\lambda}(x,y)s_{\nu^t}(-y)\Big] s_{\mu}(y)\,. \nn \eea
Since Schur functions $s_{\mu}$ form a $\Z$-basis for symmetric
functions, the coefficient of $s_{\mu}(x)$ in the above equation
should be identical. So, the proposition is proved.

As an application of the above proposition, let us take
\bea x_1 &=& q^{-\half},\quad  x_2=q^{-\frac{3}{2}}, \quad \cdots,
\quad x_{N} = q^{-N+\half}, \qquad
x_{N+1}=x_{N+2}=\cdots =0\,, \nn \\
  y_i &=& q^{-N -i +\half}\,, \quad i=1,2,3,\cdots\,, \nn \eea
which gives us $s_{\mu}(x,y) = s_{\mu}(q^{\rho})$. Then,  we can
see that
\vskip2mm

{\bf\large Corollary}:
\beq s_{\mu}(q^{-i+\half}) \equiv
s_{\mu}(q^{-\half},\cdots,q^{-N+\half}) =
\sum_{\nu}s_{\mu/\nu}(q^{\rho})s_{\nu^t}(-q^{-N} q^{\rho})\,.
\label{App}\eeq

Now, we present the skew-Schur function representation of
topological vertex, which is used in section 3. $U(\infty)$ Hopf
link invariants, $W_{\mu\nu}$, which are basic ingredients for
the construction of topological vertex, can be written by skew
Schur functions as
\bea W_{\mu\nu}(q) &=& q^{\kappa_{\nu}/2} C_{\bullet\mu\nu^t} (q)=
q^{\kappa_{\mu}/2+\kappa_{\nu}/2}\sum_{\lambda}
s_{\mu^t/\lambda}(q^{\rho})s_{\nu^t/\lambda}(q^{\rho})
=s_{\mu}(q^{\rho})s_{\nu}(q^{\mu+\rho})\,, \nn \\
W_{\bullet\mu} (q)&=& C_{\bullet\bullet\mu} (q) = s_{\mu}(q^{\rho})\,.
\label{Hopf}
\eea
Then, topological vertex is given in terms of skew Schur functions
by~\cite{Okounkov:2003sp}\cite{Eguchi:2003sj}\cite{Hollowood:2003cv}
\beq C_{\lambda\mu\nu}(q) =
C_{\mu\nu\lambda}(q)=C_{\nu\lambda\mu}(q)=
q^{\kappa_{\lambda}/2+\kappa_{\nu}/2}s_{\nu^t}(q^{\rho})
\sum_{\eta}
s_{\lambda^t/\eta}(q^{\nu+\rho})s_{\mu/\eta}(q^{\nu^t+\rho})\,.
\eeq
Note that the $\Z_2$ and $\Z_3$ symmetry of the topological vertex
implies the existence of corresponding identities  inherited from
these symmetries in skew-Schur functions.

%

\end{document}